\documentclass[a4paper, 10pt]{article}

\usepackage{amsfonts, amsmath, amssymb, amsthm, array, eucal, mathtools}
\usepackage{booktabs, array, multicol, multirow, longtable,titling,fancyhdr,titlesec}
\usepackage[runin]{abstract}
\usepackage[table]{xcolor}
\usepackage{authblk}
\usepackage[numbers,super]{natbib}
\usepackage[left=1in,top=1in,right=1in,bottom=1in,nohead]{geometry}

\title{Physics of Rheologically-Enhanced Propulsion: Different Strokes in
Generalized Stokes}
\author[1,2]{Thomas D. Montenegro-Johnson}
\author[1]{Daniel Loghin}
\author[1,2,3]{David J. Smith}
\affil[1]{School of Mathematics, University of Birmingham, Edgbaston,
Birmingham, B15 2TT, UK}
\affil[2]{Centre for Human Reproductive Science, Birmingham Women's NHS
Foundation Trust, Edgbaston, Birmingham, B15 2TG, UK}
\affil[3]{{School of Engineering \& Centre for Scientific Computing, University
of Warwick, Coventry, CV4 7AL, UK}}

\date{}

\definecolor{rulecolor}{rgb}{0.6,0.6,0.6}
\newcommand{\HRule}{\textcolor{rulecolor}{\rule[0.25cm]{\linewidth}{0.3mm}}}  

\titleformat{\section}{\large\sffamily\bfseries}{\thesection\quad}{0pt}{}[\HRule]
\titlespacing*{\section}{0cm}{0.3cm}{-0.1cm}

\titleformat{\subsection}{\normalsize\sffamily\bfseries}{\thesubsection\quad}{0pt}{}
\titlespacing*{\subsection}{0pt}{0.3cm}{0.1cm}

\pretitle{\begin{flushleft}\LARGE\sffamily\bfseries}
\posttitle{\par\end{flushleft}\vskip 0.4em}
\preauthor{\begin{flushleft}\normalsize}
\postauthor{\par\end{flushleft}\vskip 0.1em}
\predate{\begin{flushleft}\small\sc}
\postdate{\par\end{flushleft}\vskip 0.1em}

\begin{document}

\maketitle

\vspace{-1.7cm}
\begin{abstract}
Shear-thinning is an important rheological property of many biological fluids,
such as mucus, whereby the apparent viscosity of the fluid decreases with shear.
Certain microscopic swimmers have been shown to progress more rapidly through
shear-thinning fluids, but is this behavior generic to all microscopic swimmers,
and what are the physics through which shear-thinning rheology affects a
swimmer's propulsion? We examine swimmers employing prescribed stroke kinematics
in two-dimensional, inertialess Carreau fluid: shear-thinning ``Generalized
Stokes'' flow. Swimmers are modeled, using the method of femlets, by a set of
immersed, regularized forces. The equations governing the fluid dynamics are
then discretized over a body-fitted mesh and solved with the finite element
method.  We analyze the locomotion of three distinct classes of microswimmer:
(1) conceptual swimmers comprising sliding spheres employing both one- and
two-dimensional strokes, (2) slip-velocity envelope models of ciliates commonly
referred to as ``squirmers'' and (3) monoflagellate pushers, such as sperm.  We
find that morphologically identical swimmers with different strokes may swim
either faster or slower in shear-thinning fluids than in Newtonian fluids. We
explain this kinematic sensitivity by considering differences in the viscosity
of the fluid surrounding propulsive and payload elements of the swimmer, and
using this insight suggest two reciprocal sliding sphere swimmers which violate
Purcell's Scallop theorem in shear-thinning fluids. We also show that an
increased flow decay rate arising from shear-thinning rheology is associated
with a reduction in the swimming speed of slip-velocity squirmers.  For
sperm-like swimmers, a gradient of thick to thin fluid along the flagellum
alters the force it exerts upon the fluid, flattening trajectories and
increasing instantaneous swimming speed. Montenegro-Johnson et al., Phys. Fluids
25, 081903 (2013); http://dx.doi.org/10.1063/1.4818640. \textcopyright 2013
Author(s). All article content, except where otherwise noted, is licensed under
a Creative Commons Attribution 3.0 Unported License.
\end{abstract}


\section{Introduction} \label{sec:introduction}

Microscopic swimmers pervade the natural world, from bacteria and algae to the
sperm cells of animals, and the study of their swimming is pertinent to numerous
problems in medicine and industry, for example in reproductive science and
biofuel production. Microscopic self-propulsion has been a rich area of applied
mathematics for the past 60 years, motivating the development of singularity
methods such as slender body theory \citep{hancock1953self,
johnson1979flagellar} and the method of regularized stokeslets
\citep{cortez2001method}.

Because of the small length-scales of microscopic flows, viscous forces dominate
inertia. As such, there is no time dependence in the equations that govern
microscopic flow, and any periodic swimming stroke that generates net
displacement must be non-reciprocal, i.e. distinguishable from its
time-reversal. Thus, many swimming strokes that are effective at macroscopic
length-scales, such as the opening and closing of a clam shell, do not generate
progress at microscopic scales, as famously described by \citet{taylorfilm} and
\citet{purcell1977life}.

\begin{figure}[tbp]
\begin{center}
\includegraphics[scale = 0.83333]{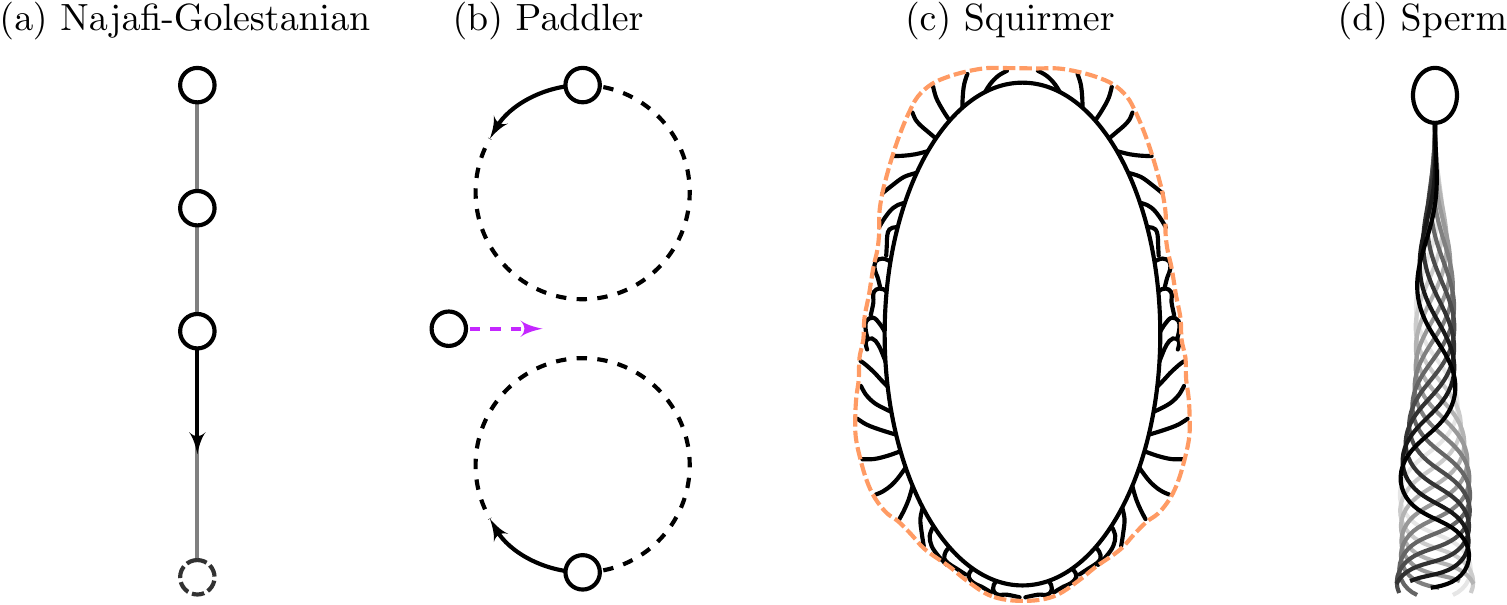}
\caption{Swimming techniques in inertialess flows that are examined in this
study. Conceptual swimmers may comprise sliding spheres that have simple
kinematics, such as (a) the collinear motion of the Najafi-Golestanian swimmer
and (b) paddling motion. These swimmers can provide insight into more complex
biological systems \citep{polotzek2012three}. (c) Ciliates beat many surface
cilia in a coordinated fashion. This is often modeled mathematically with
envelope methods, either as a small perturbation to the cell morphology
(dashed), or through a surface slip velocity.  (d) Sperm, an archetypal
``monoflagellate pusher'', propagate a bending wave down a single flagellum,
shown here in a time-lapse manner.}	
\label{fig:swimming_methods}
\end{center}
\end{figure}

Microswimmers may employ a wide variety of kinematic behaviors (figure
\ref{fig:swimming_methods}) in order to progress. For instance, sperm swim by
propagating a bending wave down a single active flagellum, whereas ciliates
``squirm'' forward through the coordinated beating of many surface cilia.
Motivated by the question of what would constitute the simplest microswimmer,
\citet{purcell1977life} considered three linked hinges undergoing periodic,
irreversible motion, which continues to inspire research, see for example
\citet{tam2007optimal,passov2012dynamics}.

A new avenue was opened for the study of simple, conceptual microswimmers by
\citet{najafi2004simple}, who showed that a swimmer comprising three sliding,
collinear spheres could progress through viscous fluid. Such models provide
insight into the physics of viscous propulsion for more complicated models
\citep{polotzek2012three, ledesma2012circle}, and may also be instructive in the
design of artificial microswimmers \citep{ogrin2008ferro}, and microfluidic
pumps. 

Many microscopic swimmers must progress through biological fluids, for example
cervical mucus \citep{lai2009micro} and bacterial extracellular slime
\citep{hall2004bacterial, verstraeten2008living}, that are suspensions of long
polymer chains. These suspended polymers endow biological fluids with complex
non-Newtonian flow properties that may impact a swimmer's ability to progress
through them. One such property that has received much recent study, both
theoretical \citep{fulford1998swimming, normand2008flapping, lauga2009life,
elfring2010two} and experimental \citep{shen2011undulatoryel}, is
viscoelasticity, whereby the fluid retains an elastic memory of its recent flow
history. In viscoelastic fluids, those swimmers exhibiting small-amplitude
oscillations are hindered \citep{lauga2007propulsion, fu2009swimming,
zhu2012self} whereas flagellates exhibiting large-amplitude waveforms can gain
propulsive advantages by timing their stroke with the fluid elastic recoil
\citep{teran2010viscoelastic}.  Additionally, reciprocal swimmers that cannot
progress in simple fluids may progress through viscoelastic fluids, in violation
of Purcell's Scallop theorem \citep{lauga2009life}.

Another important rheological property biological fluids is shear-thinning
\citep{lai2007rapid}, whereby the viscosity of the fluid decreases with flow
shear. This behavior arises from the tendency of the suspended polymers that
constitute the fluid to align locally with flow, decreasing the apparent
viscosity of the fluid. However, after early progress with modified resistive
force theories \citep{katz1980flagellar} the effects of shear-thinning on
microscopic swimming have only recently begun to be reexamined
\citep{balmforth2010microelastohydrodynamics, shen2012undulatory,
johnson2012modelling}.

\citet{johnson2012modelling} showed that the progress of two particular
swimmers, a three-sphere swimmer and a sperm-like swimmer, was enhanced by
shear-thinning rheology. This raises two questions: do all swimmers progress
more quickly in shear-thinning fluids, and what are the physical mechanisms
through which shear-thinning interacts with a swimmer's kinematics?
Furthermore, if reciprocal swimmers can progress in viscoelastic fluids, might
this also be true in shear-thinning fluids?

In this paper, we will show that other model swimmers, including the
much-studied treadmilling squirmer, may instead be hindered by shear-thinning
rheology.  We will also give quantitative and qualitative explanations of the
physical mechanisms that underlie the interactions of shear-thinning rheology
with conceptual sliding sphere swimmers, slip-velocity squirmers and sperm-like
swimmers (figure \ref{fig:swimming_methods}). Finally, based upon these
mechanisms, we suggest reciprocal sliding sphere swimmers that are able to
progress through shear-thinning fluids. We will begin by briefly describing our
mathematical and numerical modeling, which was introduced by
\citet{johnson2012modelling}.

\section{Mathematical modeling}
\label{sec:modelling}


\subsection{Fluid mechanics of microscopic swimming}

Newtonian fluid modeling has provided important insights into the mechanisms
underlying viscous propulsion. However, the need for detailed study of
non-Newtonian swimming has long been recognized
\citep{mills1978flat,katz1980new}, and experimental observations of sperm in
methylcellulose medium suggest \citep{smith2009bend} that non-Newtonian effects
may be important. We will adopt a continuum approach to modeling swimming in
biological fluids, as used in for instance \citet{lauga2007propulsion,
fu2009swimming, teran2010viscoelastic, zhu2012self}, whereby the nanoscale
structure of suspended polymers has been averaged into bulk flow properties.

At microscopic length-scales, viscous forces dominate inertia. We will examine
microscopic swimmers in inertialess generalized Stokes flow
\citep{phan2002understanding}. The equations governing the dynamics of such flow
are 
\begin{equation}
	\nabla \cdot \left(2\mu_{\mathrm{eff}}(\dot{\gamma})\boldsymbol{\varepsilon}
	(\mathbf{u})\right) - \nabla p + \mathbf{F} = 0, \quad \nabla \cdot
	\mathbf{u} = 0,	
	\label{eq:gen_stokes}
\end{equation}
for $\mathbf{u}$ the fluid velocity field, $\mu_{\mathrm{eff}}$ the effective, or
apparent, viscosity of the flow, $p$ the pressure, $\mathbf{F}$ any body
forces and $\boldsymbol{\varepsilon} (\mathbf{u}) =
(\mathbf{\nabla}\mathbf{u} + \mathbf{\nabla}\mathbf{u}^T)/2$, the strain rate
tensor.

A model of shear-thinning polymer suspensions is given by the four-parameter Carreau
constitutive law \citep{carreau1979analysis}
\begin{equation}
	\mu_{\mathrm{eff}}^{\mathrm{car}}(\dot{\gamma}) = \mu_{\infty} + (\mu_0
	- \mu_{\infty}) (1 + (\lambda \dot{\gamma} )^2 )^{(n-1)/2}, \quad 0 < n
	\leq 1,	
	\label{eq:carreau_visc}
\end{equation}
for shear rate $\dot{\gamma}  = \left(2\varepsilon_{ij}(\mathbf{u})
\varepsilon_{ij}(\mathbf{u})\right)^{1/2}$. The effective viscosity
$\mu_{\mathrm{eff}}$ of the flow decreases monotonically between a zero shear
viscosity, $\mu_0$, and an infinite shear viscosity $\mu_{\infty}$.
As the time parameter $\lambda$ increases, lower shear rates are
required to thin the fluid.

For swimmers with prescribed strokes, a characteristic velocity is given by $U =
\omega L$, where $\omega$ is the angular frequency of the swimmer's stroke and
$L$ is a characteristic length, for instance the length of the flagellum. Upon
substitution of the viscosity \eqref{eq:carreau_visc} into equations
\eqref{eq:gen_stokes} and non-dimensionalizing, we derive the dimensionless
equations,
\begin{subequations}
\begin{align}
	\hat{\nabla} \cdot \left[ 2\left( 1 + \left[ \frac{\mu_0}{\mu_\infty} - 1
	\right]\left[1 + \left(\lambda \omega
	\hat{\dot{\gamma}}\right)^2\right]^{(n-1)/2}\right)
	\hat{\boldsymbol{\varepsilon}}({\hat{\mathbf{u}}}) \right] - \hat{\nabla}
	\hat{p} + \hat{\mathbf{F}} &= 0, \\ \hat{\nabla}\cdot\hat{\mathbf{u}} &=
	0.	
	\label{eq:carreau_swimming}
\end{align}
\end{subequations}

Thus, for swimmers exhibiting prescribed beat kinematics, trajectories are
dependent only on three dimensionless quantities: the viscosity ratio
$\mu_0/\mu_\infty$, the power-law index $n$ and the shear index $\mathrm{Sh} =
\lambda \omega$ (referred to as $\mathrm{De}$ by
\citealt{johnson2012modelling}). The parameter $\mathrm{Sh}$ has the physical
interpretation of the ratio of the fluid's time parameter to the swimmer's beat
period. Newtonian flow is recovered if any of $\mu_0/\mu_\infty = 1, n = 1$ or
$\mathrm{Sh} = 0$.

This non-dimensionalization reduces the number of free parameters from four to
three. In contrast, Newtonian flow arising from prescribed boundary motion has
no free parameters. As such, the trajectories of swimmers with prescribed
kinematics in Newtonian Stokes flow exhibit no dependency on the absolute value
of the viscosity. These values only become important when considering the
magnitude of the forces on the swimmer.

\begin{figure}[tbp]
\begin{center}
\includegraphics[scale = 0.83333]{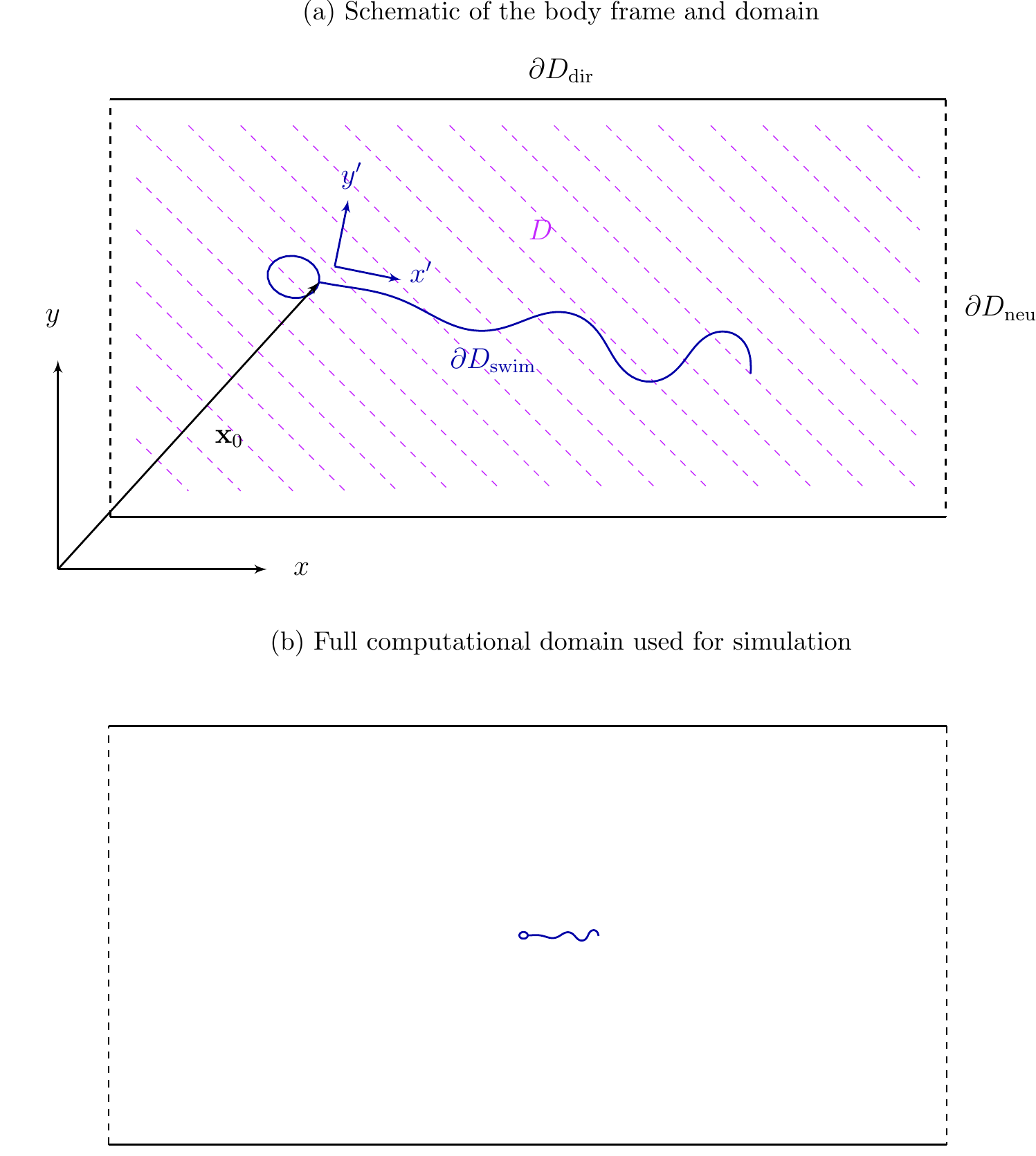}
\caption{(a) A schematic of the fluid domain $D$ containing a model human sperm
$\partial D_{\mathrm{swim}}$, showing no-slip channel walls $\partial
D_{\mathrm{dir}}$ and open boundaries $\partial D_{\mathrm{neu}}$. The
relationship between the lab frame, $(x,y)$ and the body frame,
$(x^{\prime},y^{\prime})$ is also shown, where the body frame origin
$\mathbf{x}_0$ is a fixed point on the swimmer. Femlets are distributed along
the boundary $\partial D_{\mathrm{swim}}$, shown here as a sperm head and
flagellum. (b) The full computational domain used in this study. The domain and
swimmer are shown to scale.}	
\label{fig:domain_plot}
\end{center}
\end{figure}

When prescribing the kinematics of a swimming stroke, it is convenient to employ
the swimmer's intrinsic `body frame' \citep{higdon1979hydrodynamic}, in which
its body neither rotates nor translates. The configuration and deformation of
the swimmer are specified by a mathematical function relative to the body frame,
and these are transformed into the global `lab frame' coordinates in which we
solve the governing equations.  This transformation entails use of the a priori
unknown translational velocity $\mathbf{U}$ and angular velocity
$\mathbf{\Omega}$ of the swimmer. The swimming velocities $\mathbf{U}$ and
$\mathbf{\Omega}$ result from the swimmer's body frame kinematics at any
particular time, and are constrained by the conditions that zero net force
\citep{taylor1951analysis} and torque \citep{chwang1971note} act on the swimmer.
A schematic showing the relationship between the body and lab frames is shown in
figure \ref{fig:domain_plot}a, along with the computational domain used for this
study (figure \ref{fig:domain_plot}b).

It is well-known that in two-dimensional, inertialess Newtonian flow, no
solution is possible for the flow arising from translating rigid bodies in
unbounded fluid domains. This is known as Stokes' Paradox, and arises because
the flow resulting from a point force in two dimensions diverges as $\log r$ far
from the force \citep{batchelor1967introduction}. However, the swimmers we will
model are force-free; no net forces or torques act upon them. Furthermore, since
we model swimmers in channels, the far-field decays at least as quickly as
$\mathcal{O}(1/r)$. Many cells swim close to boundaries, so that finite domain
modeling can be used to give a faithful representation of their environment. It
is highly instructive \citep{teran2010viscoelastic, crowdy2011treadmilling,
crowdy2011two} to consider two-dimensional flow models of swimming and thus we
will present results for swimmers in finite, two-dimensional domains. 

\subsection{The method of femlets}

\begin{figure}[tbp]
\begin{center}
\includegraphics[scale = 0.83333]{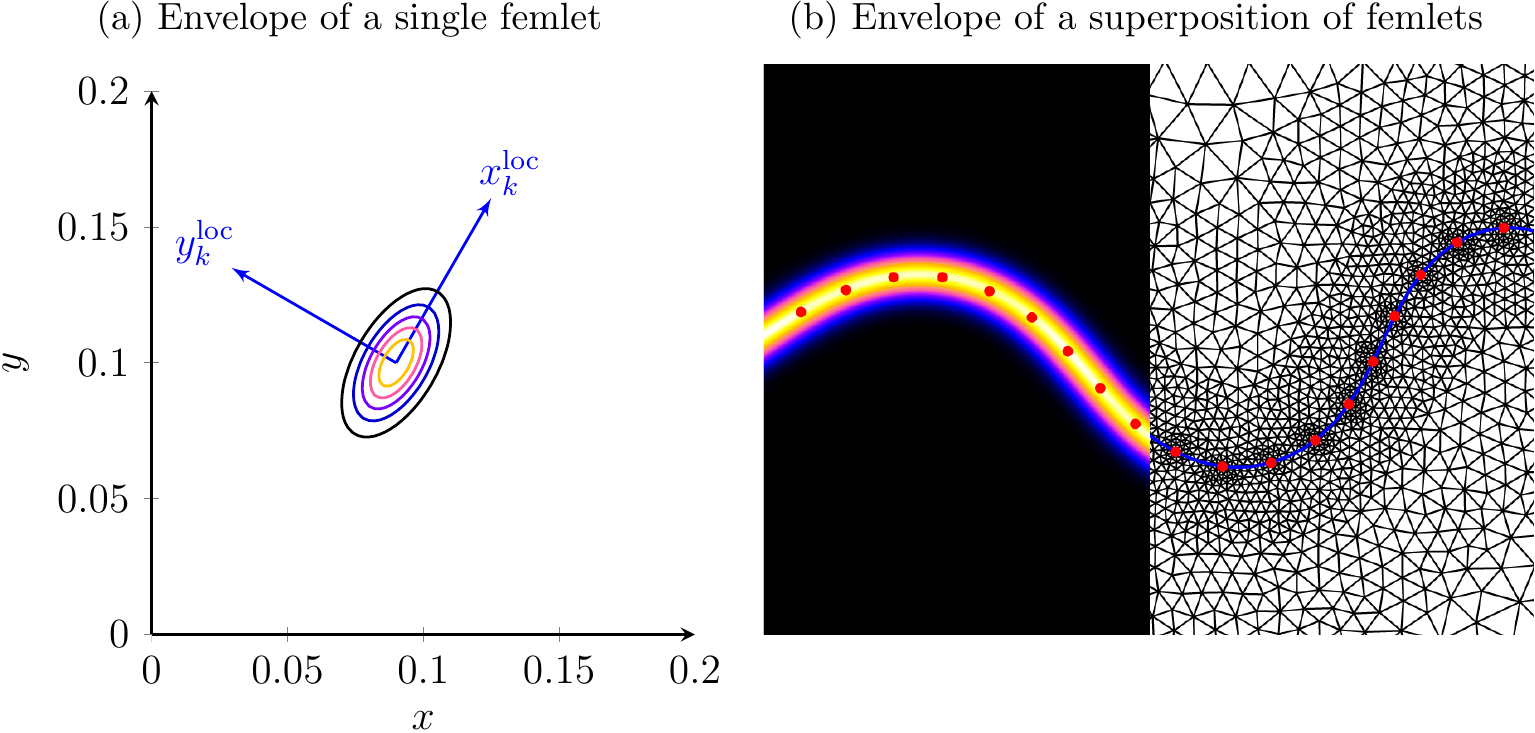}
\caption{The envelope function of the force exerted by the flagellum on the
fluid. The function is approximately zero in the black regions, and increases as
the colors lighten. (a) An example elongated femlet cut-off function, given by a
two-dimensional elongated Gaussian, oriented by a coordinate transform to align
locally with the swimmer's body.  (b) A plot showing the smooth force
distribution envelope generated by a sum of such cut-off functions when
projected on a finite element mesh; femlet centroids are marked by dots.}	
\label{fig:femlets}
\end{center}
\end{figure}

In order to solve microscopic swimming problems in fluids with shear dependent
viscosity, the method of femlets was developed by \citet{johnson2012modelling}.
Drawing inspiration from the method of regularized stokeslets
\citep{cortez2001method} and the Immersed Boundary Method \citep{peskin1972flow,
fauci1988computational}, the method of femlets represents the interaction of the
swimmer with the fluid through a set of concentrated `blob' forces of unknown
strength and direction, with spatial variation prescribed by a cut-off function
(figure \ref{fig:femlets}). While the method of regularized stokeslets reduces
the problem to finding the coefficients in a linear superposition of velocity
solutions of known form, the method of femlets proceeds by applying the finite
element method to solve for the fluid velocity field and strength and direction
of the forces simultaneously. 

For a one-dimensional filament of length $L$ and centerline parameterization
$\boldsymbol{\xi}(s,t)$, the force exerted by the filament on the fluid is given
by
\begin{equation}
\mathbf{F}(\mathbf{x},t) = \int_0^L\delta(\mathbf{x} -
\boldsymbol{\xi}(s,t))\mathbf{f}(s,t)\,\mathrm{d}s,
\label{eq:continuous_force}
\end{equation}
where $\mathbf{f}(s,t)$ is a force per unit length determined by the swimmer's
velocity. In the method of femlets, we discretize equation
\eqref{eq:continuous_force} by a set of regularized forces
\begin{equation}
\mathbf{F}(\mathbf{x},t) \approx \sum_{k=1}^{N} g^{\sigma_x,\sigma_y}\left\{ \mathbf{R}(s_k)\cdot [\mathbf{x} -
\boldsymbol{\xi}(s_k)] \right\}\mathbf{f}(s_k).
\end{equation}
The rotation $\mathbf{R}(s_k)$ is chosen such that the axis
$\mathbf{R}(s_k)\cdot [\mathbf{x} - \boldsymbol{\xi}(s_k)] =
(x_k^{\mathrm{loc}},y_k^{\mathrm{loc}})^T$ is aligned locally to the swimmer's
tangent at the location of each femlet, and $\sigma_x,\sigma$ are anisotropic
regularization parameters. For this study, we choose an elongated Gaussian
cut-off function, as in \citet{johnson2012modelling},
\begin{equation}
g^{\sigma_x,\sigma_y}\{\mathbf{x}^{\mathrm{loc}}\} =
\exp\left\{-\left[\frac{(x^{\mathrm{loc}})^2}{2\sigma^2_x} +
\frac{(y^{\mathrm{loc}})^2}{2\sigma^2_y}\right]\right\}.
\label{eq:elongated_gaussian}
\end{equation}
The regularization parameter $\sigma_x$ is chosen to give a smooth
representation along the swimmer of the force (figure
\ref{fig:femlets}b), while reducing $\sigma_y$ produces a closer
approximation to a line force (equation \eqref{eq:continuous_force}). A 
validation of the method of femlets is provided in appendix \ref{sec:appendix}.

We will model swimmers in the truncated channel $D$ shown in figure
\ref{fig:domain_plot}. On the channel walls $\partial D_{\mathrm{dir}}$ we
specify Dirichlet velocity conditions, for example the no-slip condition
$\mathbf{u}_{\mathrm{dir}} = \mathbf{0}$, and at the truncated boundary
$\partial D_{\mathrm{neu}}$ we apply the zero normal stress condition
$\boldsymbol{\sigma}\cdot \mathbf{n} = \mathbf{0}$. The swimmer $\partial
D_{\mathrm{swim}}$ is not a Dirichlet boundary, but rather a manifold of points
within $D$ on which we specify the swimmer's body frame velocity. This is where
the femlets are distributed.

For the two-dimensional problem, $2$ degrees of freedom are associated with each
femlet $k$, the lab frame force of the femlet in the $x$ and $y$ directions
$(f_{1k},f_{2k})$. This produces $2N_f$ additional scalar variables.  To
calculate the $2N_f$ force unknowns, we enforce $2N_f$ constraints in the form
of Dirichlet velocity conditions $\mathbf{u}_s$ given by the swimmer's velocity
in the body frame and applied at the location of each femlet.


\section{Results and analysis}
\label{sec:results}


\subsection{Sliding sphere swimmers}
\label{subsec:collinear_swimmers}

In the results that follow, the fluid domain is given by a channel of length
$10L$ and height $5L$, where $L$ is a characteristic length for the swimmer,
normalized here to $L = 1$ unit. To ensure the independence of the results from
the truncation length of the channel, swimmers were also tested in a channel of
length $20L$.
 
\begin{figure}[tbp]
\begin{center}
\includegraphics[scale = 0.83333]{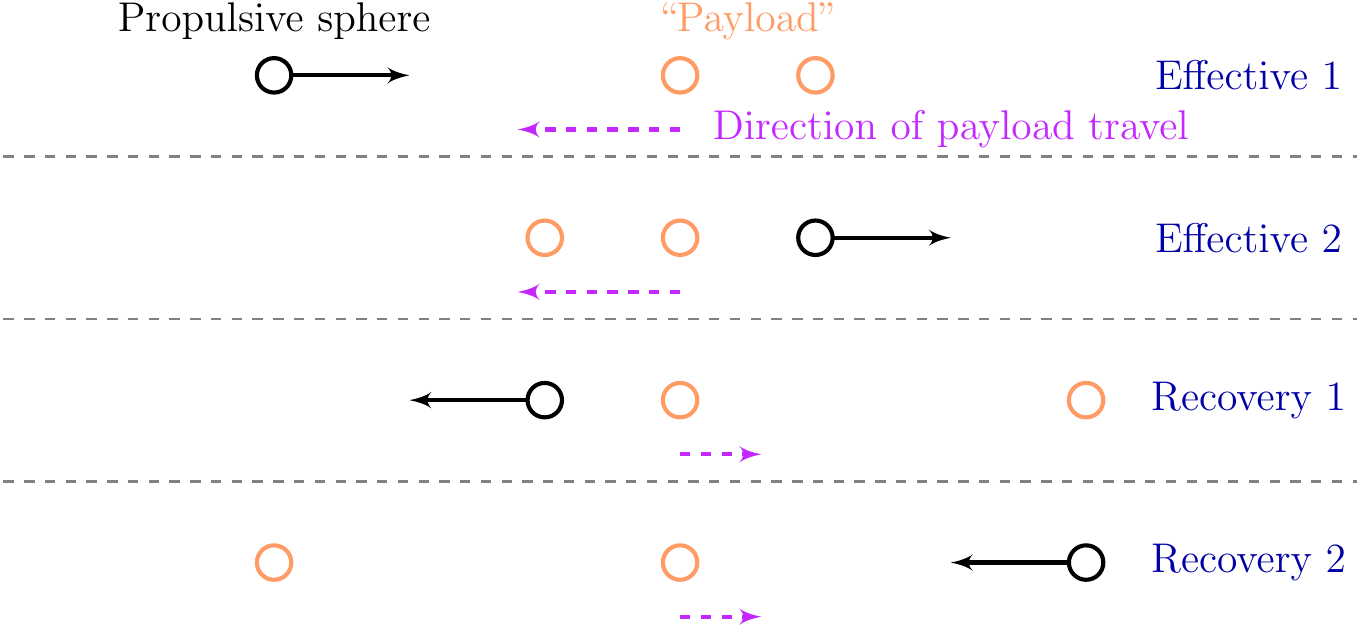}
\caption{A complete beat cycle of the Najafi-Golestanian swimmer showing the
position of the outer spheres relative to the central sphere, the direction in
which the propulsive sphere moves (solid arrow) relative to the payload, and the
direction and magnitude of swimming (dashed arrow).}
\label{fig:three_sphere_beat_pattern}
\end{center}
\end{figure}

We will begin by examining the effects of shear-thinning rheology on a class of
model viscous swimmers comprising sliding collinear spheres that oscillate out
of phase. The first such swimmer was proposed by \citet{najafi2004simple}; it is
formed of three spheres which move with the four-stage beat pattern shown in
figure \ref{fig:three_sphere_beat_pattern}. The kinematics of the beat is
divided into two ``effective'' strokes, during which the swimmer travels in the
direction of net progress, and two ``recovery'' strokes, during which the
swimmer readjusts its configuration to reinitiate an effective stroke. Whilst
performing a recovery stroke, the swimmer moves in the opposite direction to the
direction of net progress. 

We refer to the swimmer's ``progress'' as the distance it travels over an
effective stroke, ``regress'' as the distance it travels over a recovery stroke.
The swimmer's ``net progress'' is the distance travelled over an entire beat
cycle. The net progress can be seen as the sum of the distances travelled over
all effective strokes minus the sum of the distances travelled over all recovery
strokes. In other words,
\begin{equation}
\mbox{net progress} = \sum_{i=1}^{N_{\mathrm{eff}}} \mbox{progress}_i - \sum_{i=1}^{N_{\mathrm{rec}}}
\mbox{regress}_i
\end{equation}
for $N_{\mathrm{eff}}, N_{\mathrm{rec}}$ the number of effective and recovery
strokes respectively.

Figure \ref{fig:three_sphere_beat_pattern} shows that at any instant, the
swimmer can be thought of as comprising a propulsive element and a drag-inducing
``payload'' element.  By force balance, leftward relative motion of an outer
sphere results in rightward motion of the remaining spheres through the fluid,
and vice-versa.  The principle underlying the propulsion of collinear sphere
swimmers is that the total drag on the two payload spheres is reduced if they
are brought closer together. Thus, the swimmer shown in figure
\ref{fig:three_sphere_beat_pattern} will exhibit overall leftward progress.

\citet{johnson2012modelling} found that a version of the Najafi-Golestanian
swimmer with smoothed kinematics progressed more rapidly through shear-thinning
fluid. However, the physics behind this enhanced progression were not apparent.
We will now consider the simpler original Najafi-Golestanian swimmer, for which
the outer spheres move at constant speed during each portion of the four-stage
beat cycle shown in figure \ref{fig:three_sphere_beat_pattern}.  The body frame
positions of the three spheres $i = 1,2,3$ are given as a function of time $t$
in table \ref{tab:three_sphere}, where $d = 0.5, a = 0.25$ in our model.

\begin{table}
\begin{center}
\begin{tabular}{@{}lcccr@{}} \toprule
\multicolumn{5}{c}{Najafi-Golestanian swimmer} \\
Stroke  & $x_1$			& $x_2$ & $x_3$ 	    & time $t$ \\ \midrule
Eff $1$ & $-(d+a) + 8at$       	& $0$   & $d-a$             & $[0,1/4)$  \\         
Eff $2$ & $-(d-a)$             	& $0$   & $d-a + 8a(t-1/4)$ & $[1/4,1/2)$  \\
Rec $1$ & $-(d-a) - 8a(t-1/2)$ 	& $0$   & $d+a$             & $[1/2,3/4)$  \\
Rec $2$ & $-(d+a)$             	& $0$   & $d+a - 8a(t-3/4)$ & $[3/4,1)$ \\ \bottomrule
\end{tabular}
\end{center}
\caption{The body frame positions of the three spheres of the Najafi-Golestanian
swimmer we will model, for $d = 0.5, a = 0.25$, over each portion of its beat cycle.}
\label{tab:three_sphere}
\end{table}

\begin{figure}[tbp]
\begin{center}
\includegraphics[scale = 0.83333]{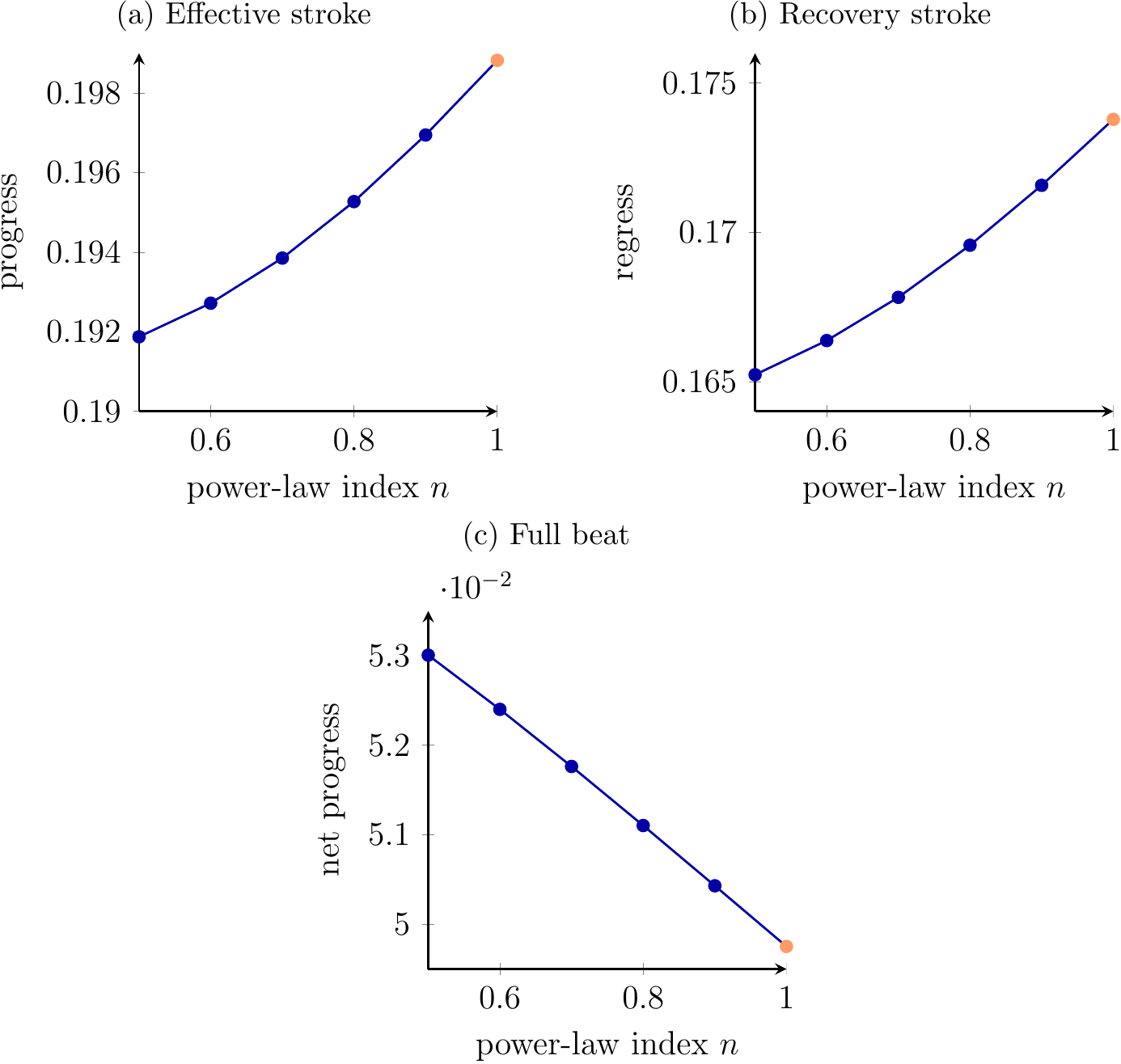}
\caption{The effects of shear-thinning on the Najafi-Golestanian swimmer with
the four-stage beat pattern given in table \ref{tab:three_sphere}.  (a) The
progress during each effective stroke and (b) the regress during each recovery
stroke as functions of the power-law index $n$. Since the decrease in regress is
greater for $n < 1$, the overall effect of shear-thinning is an increase in net
progress as $n$ decreases (c). In each panel, the case corresponding to
Newtonian fluid is marked in lighter gray (orange online).}	
\label{fig:three_sphere_vs_n}
\end{center}
\end{figure}

\begin{figure}[tbp]
\begin{center}
\includegraphics[scale = 0.83333]{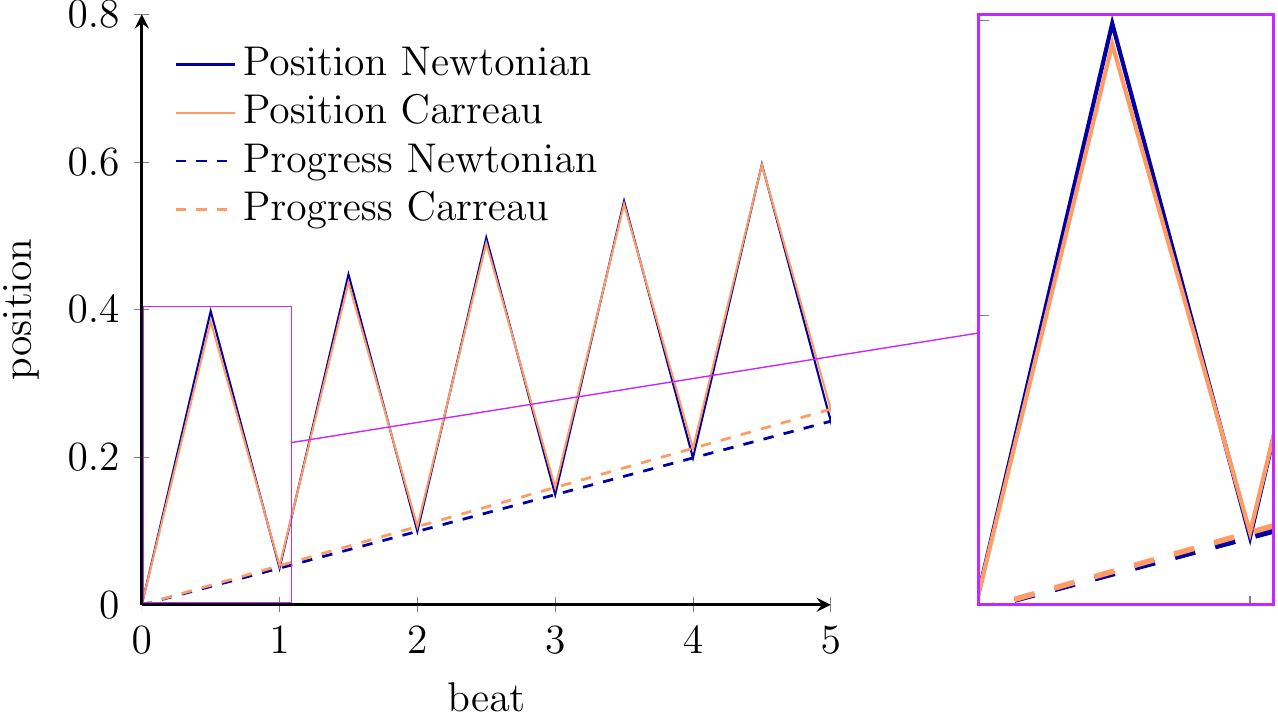}
\caption{Simulation results of the position of the Najafi-Golestanian swimmer
over five beat cycles, demonstrating how decreasing the instantaneous swimming
speed at all times in shear-thinning fluid can lead to an increase in overall
progress, provided swimming speed is decreased more during the recovery stroke.
The rheological parameters of the Carreau fluid are $\mu_0/\mu_\infty = 2, n =
0.5$ and $\mathrm{Sh} = 1$.}	
\label{fig:shear_thinning_eff_rec_demo}
\end{center}
\end{figure}

Figure \ref{fig:three_sphere_vs_n} shows the effects of shear-thinning rheology
upon the Najafi-Golestanian swimmer for varying power-law index $n$.  As $n$ is
decreased from the Newtonian case $n=1$, the swimmer's progress over its
effective strokes (figure \ref{fig:three_sphere_vs_n}a) and regress over
recovery strokes (figure \ref{fig:three_sphere_vs_n}b) are both decreased. At
all moments during its beat cycle, the swimmer swims more slowly in
shear-thinning fluid. This effect is slight: for $n = 0.5$, the swimmer's speed
is approximately $3 \%$ lower during the effective strokes and $5 \%$ lower
during the recovery strokes than for $n = 1$ (Newtonian fluid). However, since
swimming velocity is reduced more during the recovery strokes, the result is in
fact an increase in net progress, shown in figure \ref{fig:three_sphere_vs_n}c.
This behavior is demonstrated in figure \ref{fig:shear_thinning_eff_rec_demo},
which shows the position of the swimmer over five complete beat cycles in
Newtonian and shear-thinning fluid.

The swimmer's progress and regress are reduced by shear-thinning, but regress is
reduced more and hence overall progress is increased. But what is responsible
for this decrease in instantaneous swimming speed, and why is this effect
enhanced during the recovery stroke? 

\begin{figure}[tbp]
\begin{center}
\includegraphics[scale = 0.83333]{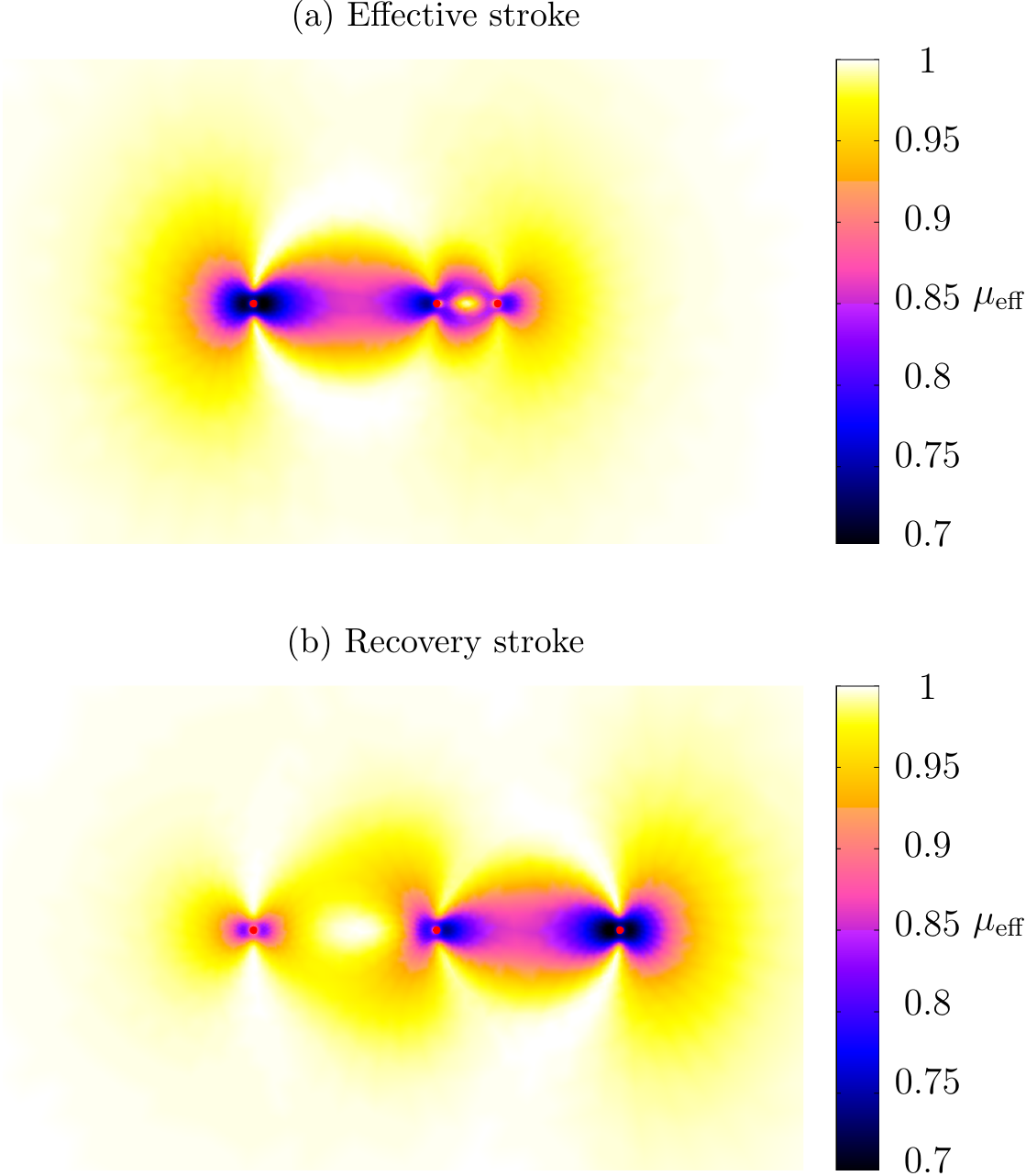}
\caption{The effective viscosity of Carreau fluid, normalized to $\mu_0 = 1$,
surrounding the Najafi-Golestanian swimmer (table \ref{tab:three_sphere}) at (a)
the start of effective stroke $1$ and (b) the start of recovery stroke $2$ for
$\mu_0/\mu_\infty = 2, n = 0.5$ and $\mathrm{Sh} = 1$. The fluid around the
propulsive sphere is thinner than that around the payload.}
\label{fig:golestanian_viscosity}
\end{center}
\end{figure}

Figure \ref{fig:golestanian_viscosity} shows the effective viscosity of the
fluid surrounding the swimmer at time $t = 0$ for rheological parameters
$\mu_0/\mu_\infty = 2, n = 0.5$ and $\mathrm{Sh} = 1$. The effective viscosity
of the fluid surrounding the propulsive sphere is significantly lower than that
surrounding the payload. In the lab frame, the propulsive sphere moves more
quickly than the payload, thereby thinning the surrounding fluid to a greater
extent.

The drag on a sphere moving in inertialess Newtonian fluid is proportional to
the viscosity of the fluid. Whilst Carreau fluid is non-Newtonian, this
observation is key to understanding the effects of shear-thinning rheology.
If fluid is relatively thicker around the payload spheres, the resistance
coefficient of those spheres will be relatively higher than that of the
propulsive sphere. Thus, the instantaneous velocity of the swimmer will be
reduced.

\begin{figure}[tbp]
\begin{center}
\includegraphics[scale = 0.83333]{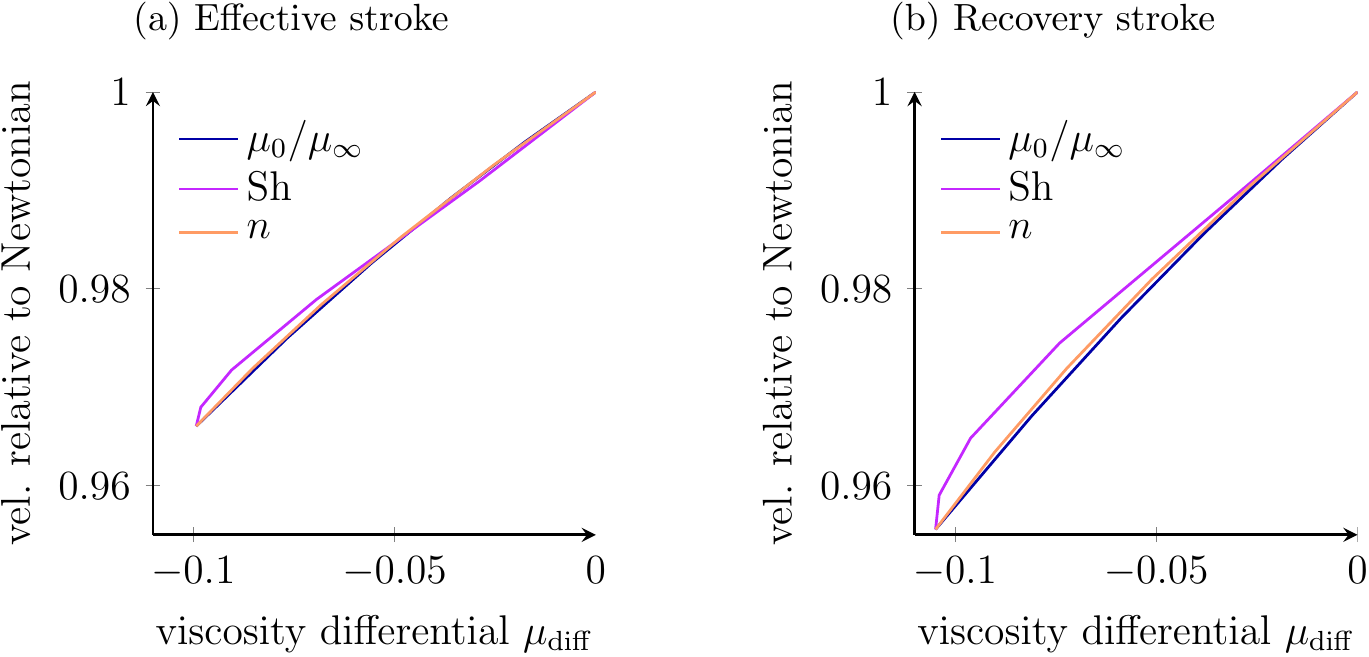}
\caption{The velocity relative to the Newtonian case of the Najafi-Golestanian
swimmer when initiating an effective stroke (a) and a recovery stroke (b) as a
function of the viscosity differential $\mu_{\mathrm{diff}}$. The velocity has
been calculated while varying the three rheological parameters of Carreau flow
for $n =0.5, \mu_0/\mu_\infty \in [1, 2], \mathrm{Sh} = 0.5$ (dark gray, blue
online), $n = 0.5, \mu_0/\mu_\infty = 2, \mathrm{Sh} \in [0, 0.5]$ (medium gray,
magenta online) and $n \in [0.5, 1], \mu_0/\mu_\infty = 2, \mathrm{Sh} = 0.5$
(light gray, orange online). This figure demonstrates an apparent
proportionality between the velocity and the viscosity differential, and that
the viscosity differential is enhanced during the recovery stroke.}	
\label{fig:three_sphere_visc_diff}
\end{center}
\end{figure}

We examine this effect by calculating the average viscosity of the flow at
points on a small circle, of radius $\epsilon$ say, surrounding each sphere $i$
centered at $(x_i,y_i)$
\begin{equation}
\bar{\mu}_i = \bar{\mu}_{\mathrm{eff}}(\dot{\gamma}(\mathbf{u}(r_i))),
\end{equation}
for $r_i$ coordinates $(x,y)$ such that $(x - x_i)^2 + (y - y_i)^2 =
\epsilon^2$. The average for each sphere is calculated from $20$ azimuthal
coordinates. We then split the set of viscosities into the viscosities of the
fluid surrounding propulsive $\mu^{\mathrm{prop}}_i$ and drag-inducing payload
$\mu^{\mathrm{drag}}_i$ spheres. We then calculate the ``viscosity differential''
\begin{equation}
\mu_{\mathrm{diff}} = \frac{1}{N_\mathrm{prop}}\sum_{i=1}^{N_\mathrm{prop}} \mu^{\mathrm{prop}}_i - 
\frac{1}{N_\mathrm{drag}}\sum_{i=1}^{N_\mathrm{drag}} \mu^{\mathrm{drag}}_i,
\label{eq:visc_diff}
\end{equation}
for $N_\mathrm{prop}$ and $N_\mathrm{drag}$ the number of propulsive and
drag-inducing spheres respectively. For the Najafi-Golestanian swimmer,
$N_\mathrm{prop} = 1$ and $N_\mathrm{drag} = 2$, and the propulsive and payload
spheres change according to the portion of the beat cycle, as demonstrated in
figure \ref{fig:three_sphere_beat_pattern}. The decrease in the Najafi-Golestanian
swimmer's instantaneous velocity is shown as a function of the viscosity
differential \eqref{eq:visc_diff} in figure \ref{fig:three_sphere_visc_diff}. 

At time $t=0$, the swimmer initiates an effective stroke. The velocity of the
swimmer at $t=0$, relative to the Newtonian case, is shown as a function of
$\mu_{\mathrm{diff}}$ in figure \ref{fig:three_sphere_visc_diff}a, for
varying $n$ (light gray, orange online), $\mu_0/\mu_\infty$ (dark gray, blue
online) and $\mathrm{Sh}$ (medium gray, magenta online). This figure shows that
the result of varying these parameters is approximately equivalent with respect
to the viscosity differential. Furthermore, figure
\ref{fig:three_sphere_visc_diff}a demonstrates that the reduction in velocity
arising from shear-thinning rheology is approximately proportional to the
viscosity differential. This proportionality is to be expected, because the the
drag coefficients of the spheres are approximately proportional to the viscosity
of the fluid surrounding them.

However, the coefficient of proportionality between the relative instantaneous
velocity and the viscosity differential is greater during the recovery stroke
(figure \ref{fig:three_sphere_visc_diff}b). This increase entails that the
velocity is decreased more during the recovery stroke, and must arise not from
the viscosity at the surface of the spheres, but in some way from the rate at
which the viscosity field increases away from each sphere.

These results raise three interesting questions: (1) is the viscosity
differential always negative, reducing instantaneous velocity, for three-sphere
swimmers, (2) will the coefficient of proportionality between the instantaneous
velocity and the viscosity differential always be greater during the recovery
stroke and (3) how does the rate at which viscosity increases away from the
swimmer affect progress? To answer these questions, we will first consider a
morphologically identical three-sphere swimmer with different beat kinematics.
 

\subsection{A three-sphere ``paddler''}

\begin{figure}[tbp]
\begin{center}
\includegraphics[scale = 0.83333]{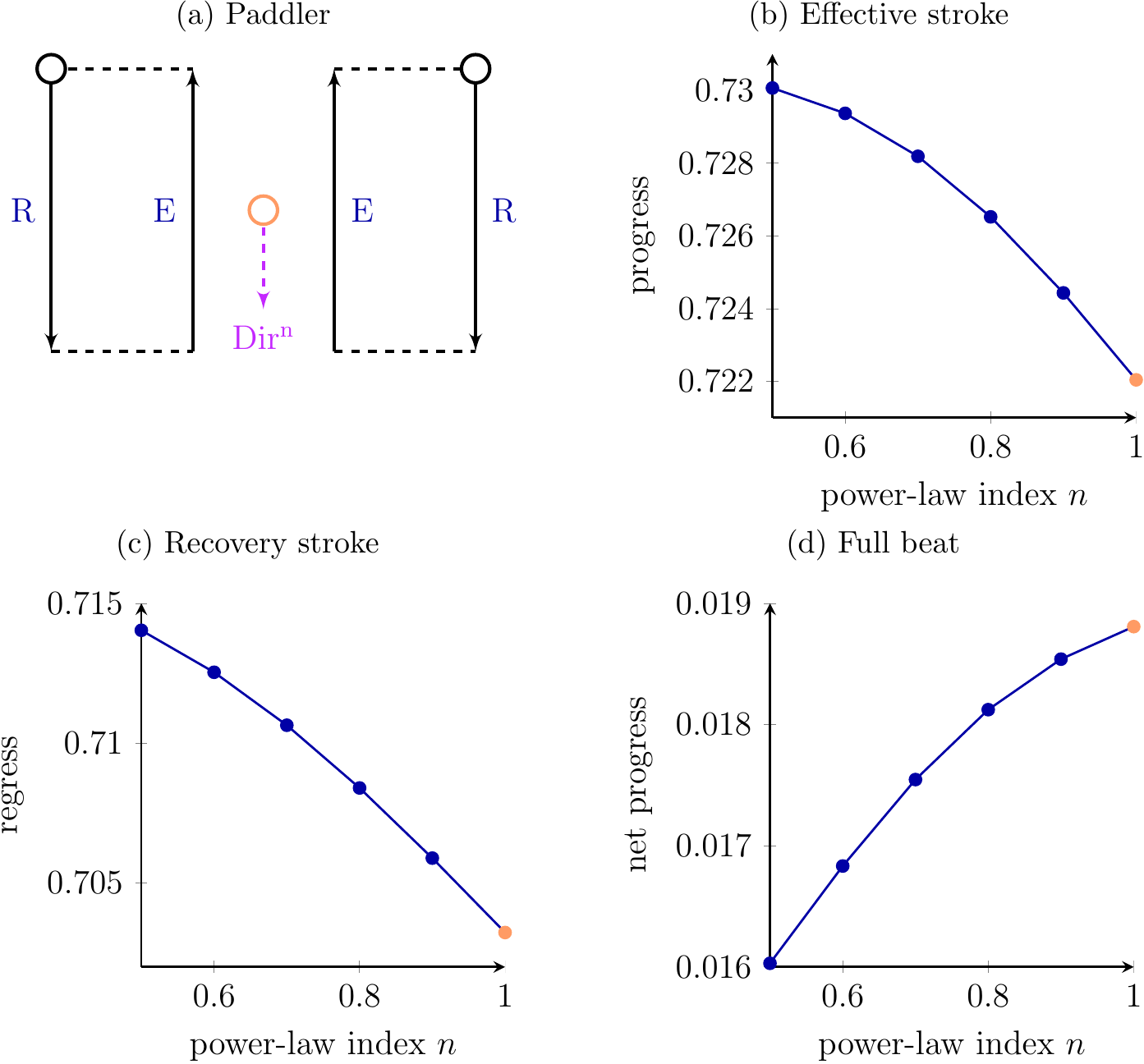}
\caption{The effects of shear-thinning on the paddler (a) with the two-stage
beat pattern given in table \ref{tab:three_sphere_chlamy}. During the portions
of the beat represented by the dashed black lines, the swimmer does not progress
and as such they are not considered here. The dashed arrow shows the swimming
direction. (b) The progress during the effective stroke and (c) the regress
during the recovery stroke as functions of the power-law index $n$. The greater
increase in regress results in a decrease in net progress with shear-thinning
rheology, (d). In each panel, the case corresponding to Newtonian fluid is
marked in lighter gray (orange online).}
\label{fig:three_sphere_chlamy_vs_n}
\end{center}
\end{figure}

\citet{drescher2010} showed that the far-field flow induced by the the
biflagellate green alga \textit{Chlamydomonas Reinhardtii} may be approximated
by three stokeslets: two outer stokeslets exerted a backwards force,
representing the flagella, balanced by a central stokeslet, representing the
cell body. Inspired by this approximation, one could consider a paddling
three-sphere swimmer \citep{polotzek2012three} exhibiting the kinematics shown
in figure \ref{fig:three_sphere_chlamy_vs_n}a.

The central sphere is stationary in the body frame, and represents the swimmer's
body, or payload. The two outer spheres move along closed, non-intersecting
curves in the same plane as the body, such that these curves are a mirror image
of one another. The behavior of this swimmer in Newtonian fluid was analyzed by
\citet{polotzek2012three}; it was shown that the direction the swimmer travels
is dependent upon the loci of the outer swimming spheres.

We will consider a swimmer for which the swimming spheres move along rectangles,
centered in line with body sphere.  The effective stroke occurs when the outer
spheres are nearer the body, so that the swimmer shown in figure
\ref{fig:three_sphere_chlamy_vs_n}a will generate a net displacement downwards.
Since no net motion of the swimmer occurs whilst the swimming arms are moving
directly towards or away from one another, we may consider only the two parts of
the stroke given in table \ref{tab:three_sphere_chlamy}.

\begin{table}
\begin{center}
\begin{tabular}{@{}lcccr@{}} \toprule
\multicolumn{5}{c}{Three sphere paddler}\\
Stroke & $(x_1, y_1)$ 		      	     & $(x_2, y_2)$ & $(x_3, y_3)$ 	                    & time $t$   \\ \midrule
Rec    & $d - 4dt, y_{\mathrm{rec}}$   	     & $0,0$        & $d - 4dt, -y_{\mathrm{rec}}$          & $[0,1/2)$  \\         
Eff    & $-d + 4d(t - 1/2), y_{\mathrm{eff}}$ & $0,0$        & $-d + 4d(t - 1/2), -y_{\mathrm{eff}}$ & $[1/2,1)$  \\ \bottomrule
\end{tabular}
\end{center}
\caption{The body frame positions of the three spheres for the paddling swimmer
over the effective and recovery stroke, where in our model $d = 0.5,
y_{\mathrm{rec}} = 0.75$ and $ y_{\mathrm{eff}} = 0.25$.}
\label{tab:three_sphere_chlamy}
\end{table}

\begin{figure}[tbp]
\begin{center}
\includegraphics[scale = 0.83333]{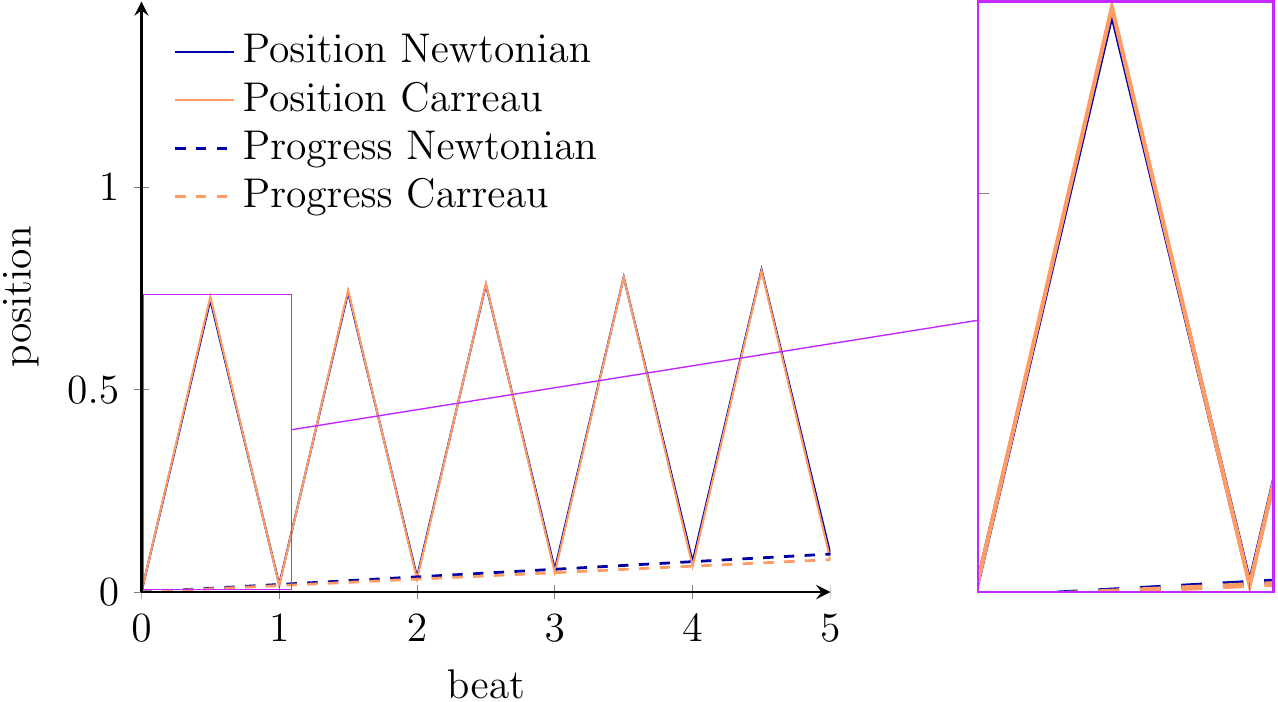}
\caption{Simulation results of the position of the paddler over five beat
cycles, demonstrating how increasing the instantaneous swimming speed at all
times in shear-thinning fluid can lead to an decrease in net progress, provided
swimming speed is decreased more during the recovery stroke. The observed effect
is exactly opposite to that of the Najafi-Golestanian swimmer, summarized in
figure \ref{fig:shear_thinning_eff_rec_demo}.  The rheological parameters of the
Carreau fluid are $\mu_0/\mu_\infty = 2, n = 0.5$ and $\mathrm{Sh} = 1$.}	
\label{fig:paddler_shear_thinning_eff_rec_demo}
\end{center}
\end{figure}

For $d = 0.5, y_{\mathrm{rec}} = 0.75$ and $ y_{\mathrm{eff}} = 0.25$, figures
\ref{fig:three_sphere_chlamy_vs_n}b and \ref{fig:three_sphere_chlamy_vs_n}c show
the swimmer's progress and regress over its effective and recovery strokes
respectively. In contrast to the Najafi-Golestanian swimmer considered above,
shear-thinning increases the instantaneous swimming speed of this paddler.
Progress is increased by around $1 \%$, and regress by around $2 \%$. The result
is a decrease in net progress (figure \ref{fig:three_sphere_chlamy_vs_n}d).
Thus, despite swimming more quickly at all times, this swimmer is hindered by
shear-thinning flow. This behavior is demonstrated in figure
\ref{fig:paddler_shear_thinning_eff_rec_demo}, which shows the position of the
swimmer over five complete beat cycles in Newtonian and shear-thinning fluid.

As with the Najafi-Golestanian swimmer, the effect of shear-thinning is small
for the parameters considered. However, these effects are sensitive to
kinematics. The Najafi-Golestanian swimmer and the paddler both comprise three
sliding spheres, but through their kinematics they are affected by
shear-thinning in opposite manners.

\begin{figure}[tbp]
\begin{center}
\includegraphics[scale = 0.83333]{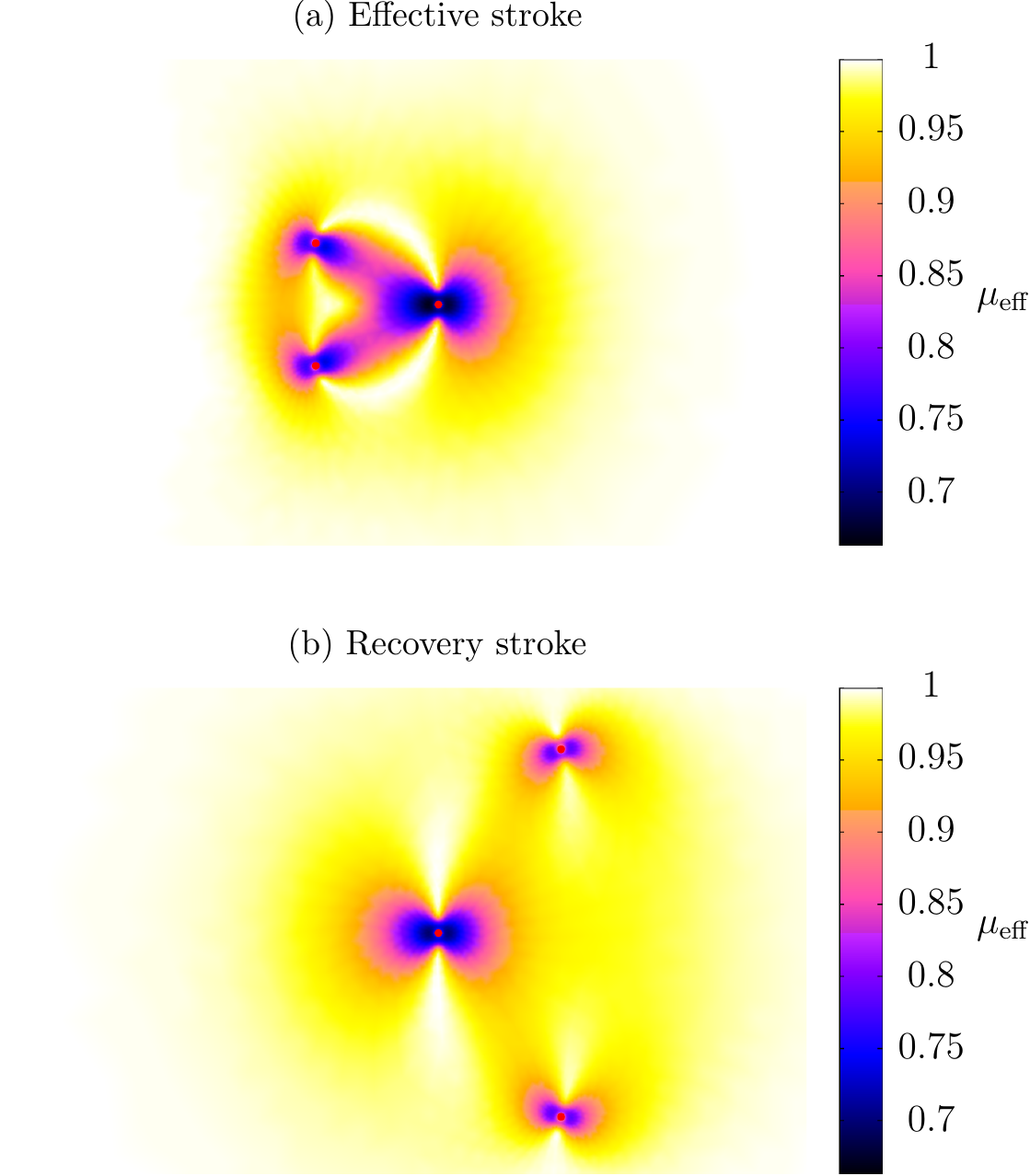}
\caption{The effective viscosity of Carreau fluid, normalized to $\mu_0 = 1$,
surrounding the paddler (table \ref{tab:three_sphere_chlamy}) at (a) the start
of the effective stroke and (b) the start of the recovery stroke for
$\mu_0/\mu_\infty = 2, n = 0.5$ and $\mathrm{Sh} = 1$. The fluid around the
propulsive sphere is thinner than that around the payload.}
\label{fig:paddler_viscosity}
\end{center}
\end{figure}

To balance the forces induced by the two propulsive spheres, the lab frame
velocity of the drag-inducing sphere is greater than the lab frame velocity of
the propulsive spheres. Thus in shear-thinning flow, fluid will be relatively
thinner around the drag-inducing sphere than around the propulsive spheres
(figure \ref{fig:paddler_viscosity}).  Accordingly, the viscosity differential
for this swimmer is positive, in contradistinction to the Najafi-Golestanian
swimmer above, and thus the swimmer's instantaneous velocity is increased by
shear-thinning rheology. But why is this effect enhanced during the recovery
stroke when the spheres are further apart?

\begin{figure}[tbp]
\begin{center}
\includegraphics[scale = 0.83333]{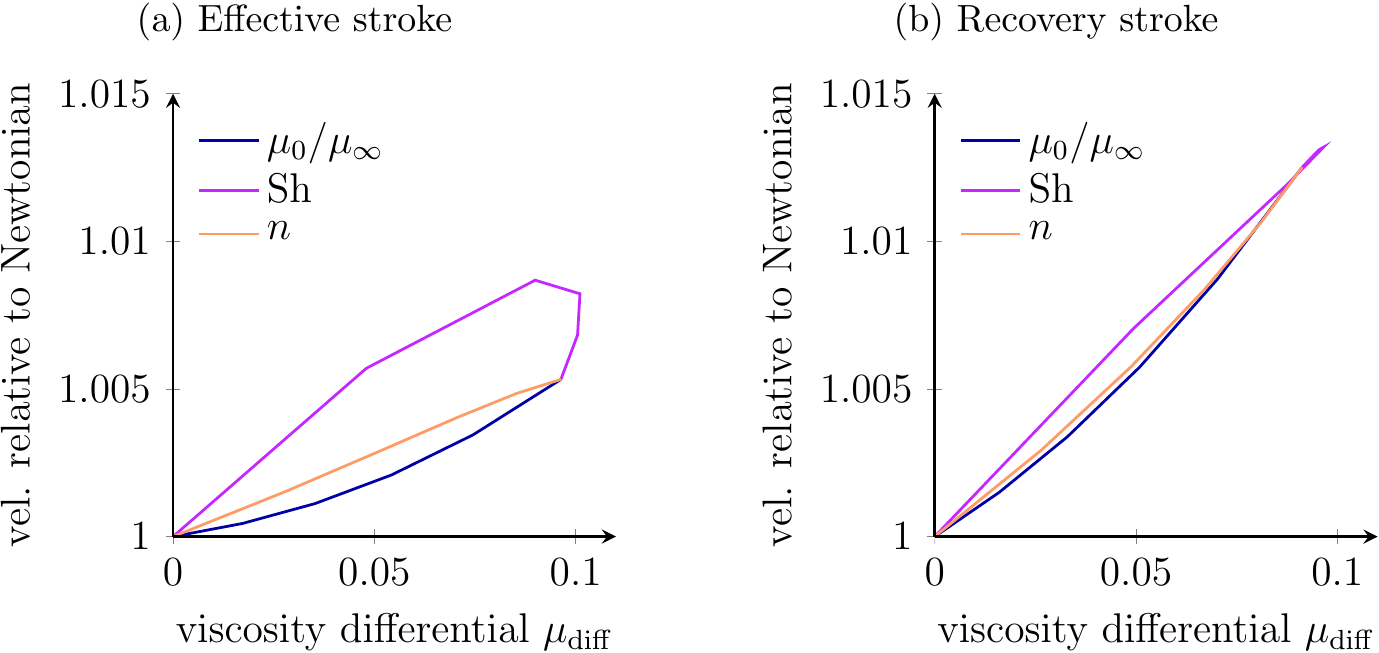}
\caption{The velocity relative to the Newtonian case of the paddler at the
commencement of (a) an effective stroke and (b) a recovery stroke as functions
of the viscosity differential $\mu_{\mathrm{diff}}$. The velocity has been
calculated while varying the three rheological parameters of Carreau flow for $n
=0.5, \mu_0/\mu_\infty \in [1, 2], \mathrm{Sh} = 0.5$ (dark gray, blue online),
$n = 0.5, \mu_0/\mu_\infty = 2, \mathrm{Sh} \in [0, 0.5]$ (medium gray, magenta
online) and $n \in [0.5, 1], \mu_0/\mu_\infty = 2, \mathrm{Sh} = 0.5$ (light
gray, orange online). During the recovery stroke (b), spheres are far apart and
there is approximate proportionality between the increase in velocity and the
viscosity differential. During the effective stroke (a), however, interactions
between the viscosity fields of the spheres reduce the effect of the viscosity
differential. For low values of $\mathrm{Sh}$ (medium gray, magenta online),
more shear is required to thin the flow. Thus, proportionality between velocity
increase and viscosity differential is maintained with the same constant for
effective and recovery strokes due to decreased viscosity field interactions.}
\label{fig:paddler_visc_diff}
\end{center}
\end{figure}

Figure \ref{fig:paddler_visc_diff} shows the velocity of the swimmer relative to
the Newtonian case as a function of the viscosity differential at a moment
during an effective stroke (figure \ref{fig:paddler_visc_diff}a) and a
recovery stroke (figure \ref{fig:paddler_visc_diff}b). During the recovery
stroke, the velocity relative to the Newtonian case is again approximately
proportional to the viscosity differential. The constant of proportionality is
approximately half that for the Najafi-Golestanian swimmer (figure
\ref{fig:three_sphere_visc_diff}), which may be because there are twice as many
propulsive elements.

However, this proportionality fails during the effective stroke, when the
spheres are close to one another. Each sphere thins a significant region of
fluid, and these regions overlap substantially, decreasing the effect of the
viscosity differential. This decrease is apparent when considering the
shear-index data in figure \ref{fig:paddler_visc_diff}a.  For low values
of $\mathrm{Sh}$, high shear is required to thin the flow. Thus, the viscosity
fields generated by the spheres that comprise the swimmer do not interact, and
the proportionality between the viscosity index and the increase in velocity is
equal to that during the recovery stroke (figure
\ref{fig:paddler_visc_diff}b), for which the spheres are further apart.
When the value of $\mathrm{Sh}$ increases past a critical value, despite
increases in the viscosity differential velocity is in fact decreased. After a
further critical value, the viscosity differential is in fact decreased by
increasing $\mathrm{Sh}$. 

The envelope of thinned fluid surrounding the swimmer during the
effective stroke inhibits its progress. Increasing the shear index past the
optimum increases the size of this envelope, further hindering swimming. This
result is consistent with the existence of an optimum value of $\mathrm{Sh}$ for
the progress of the Najafi-Golestanian swimmer considered by
\citet{johnson2012modelling}.

Thus, in the limit of large separation between spheres, the envelopes of thinned
fluid surrounding each sphere do not interact, and instantaneous velocity is
approximately proportional to the viscosity differential. If spheres are close
enough to generate an envelope of thinned fluid surrounding the whole swimmer,
that envelope hinders swimming, reducing the constant of proportionality between
swimming velocity and the viscosity differential. To examine the effects of the
envelope of thinned fluid further, we will now consider squirming models of
ciliates.


\subsection{Slip velocity squirmers}
\label{subsec:squirming_swimmers}

\begin{figure}[tbp]
\begin{center}
\includegraphics[scale = 0.83333]{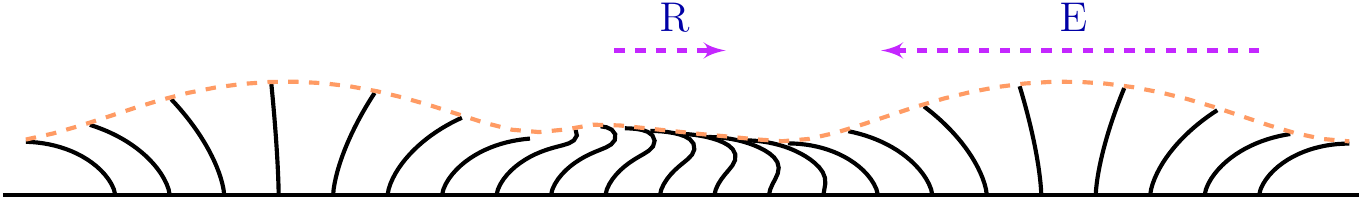}
\caption{A schematic of a ciliated surface. Cilia beat with an
effective-recovery stroke pattern, marked with E and R respectively, pushing
fluid locally in the direction shown. The cilia are activated in a coordinated,
metachronal fashion. The envelope of this motion is given by the dashed
green line.}
\label{fig:metachronal_wave}
\end{center}
\end{figure}

Much like sphere swimmers, cilia utilized for locomotion typically beat with an
asymmetric effective-recovery stroke pattern \citep{blake1974mechanics}. They
perform an effective stroke when fully extended, moving through the fluid
perpendicular to their centerline, and then recover by moving tangentially to
their centerline (figure \ref{fig:metachronal_wave}). 

Ciliated swimmers generally express a large number of cilia which beat with a
phase difference between neighbors \citep{childress1981mechanics}. Examples are
the protozoa \textit{Opalina} and \textit{Paramecium} \citep{brennen1977fluid}
and the alga \textit{Volvox Carteri}. This type of swimming motivates `envelope'
modeling approaches \citep{blake1971spherical} whereby the array of cilia are
represented by either a slip velocity condition on the cell surface, or by
small `squirming' deformations of the cell body \citep{ishikawa2006hydrodynamic,
lin2011stirring}.

We will analyze a model swimmer with a time independent stroke, the effects of
coordinated ciliary beating being time averaged over a beat as a constant slip
velocity. The tangential slip velocity is typically decomposed into `swimming
modes' of spherical harmonics \citep{michelin2011optimal}
\begin{equation}
	u_\theta (\theta) = \sum _{n=1}^\infty \alpha _n K_n (\cos \theta),
\end{equation}
for
\begin{equation}
	K_n (\cos \theta) = \frac{(2n + 1)\sin \theta}{n(n+1)} L_n ^{\prime}
	(\cos \theta ),
\end{equation}
with $L_n (\cos \theta )$ the $n$-th Legendre polynomial. Thus, slip velocity
squirmers are characterized by the coefficients $\alpha _n$ of the modes of
their swimming.

\begin{figure}[tbp]
\begin{center}
\includegraphics[scale = 0.83333]{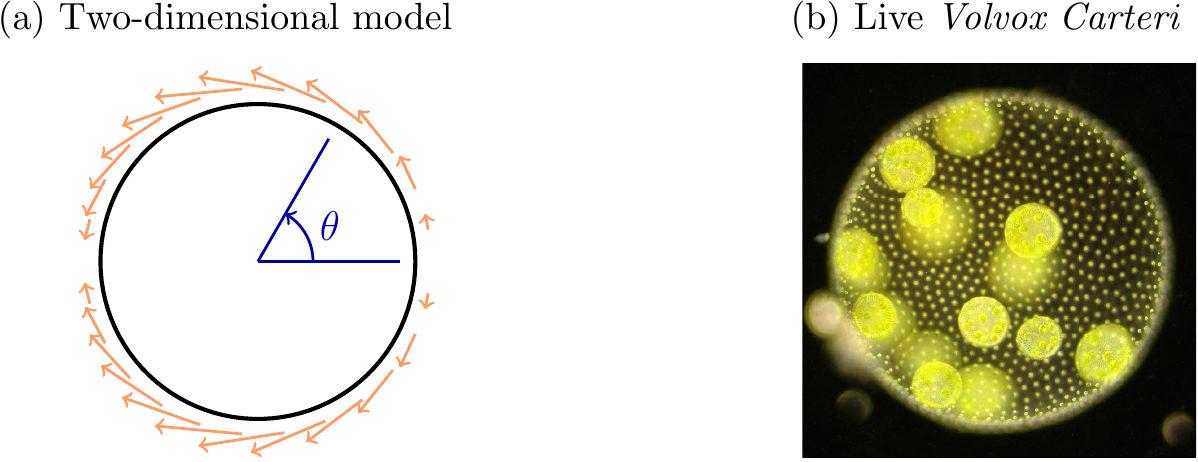}
\caption{(a) A schematic of a two-dimensional treadmilling squirmer, along with
(b) a micrograph of a \textit{Volvox Carteri} colony, showing surface cilia that
beat in a coordinated fashion to propel the colony forwards. This cell also
shows a number of characteristic `daughter' colonies within it. Image taken by
Prof. Raymond E.  Goldstein, University of Cambridge.}
\label{fig:squirmer_volvox}
\end{center}
\end{figure}

The simplest two-dimensional squirmer has a single mode, i.e. $\alpha _n = 0$
for all $n \geq 2$. This `treadmilling' squirmer has a radius $r = L/2$ and
generates a time independent tangential slip velocity in the body frame of
\begin{equation}
	u_\theta = (1/2)\sin \theta \ \mbox{on} \ r = L/2.
	\label{eq:squirmer_param}
\end{equation}
A treadmilling squirmer is shown alongside an image of \textit{Volvox Carteri},
in figure \ref{fig:squirmer_volvox}. Since swimmer kinematics and the fluid
domain are symmetric about the line $y = 0$, the squirmer swims purely in the
positive $x$ direction.

\begin{figure}[tbp]
\begin{center}
\includegraphics[scale = 0.83333]{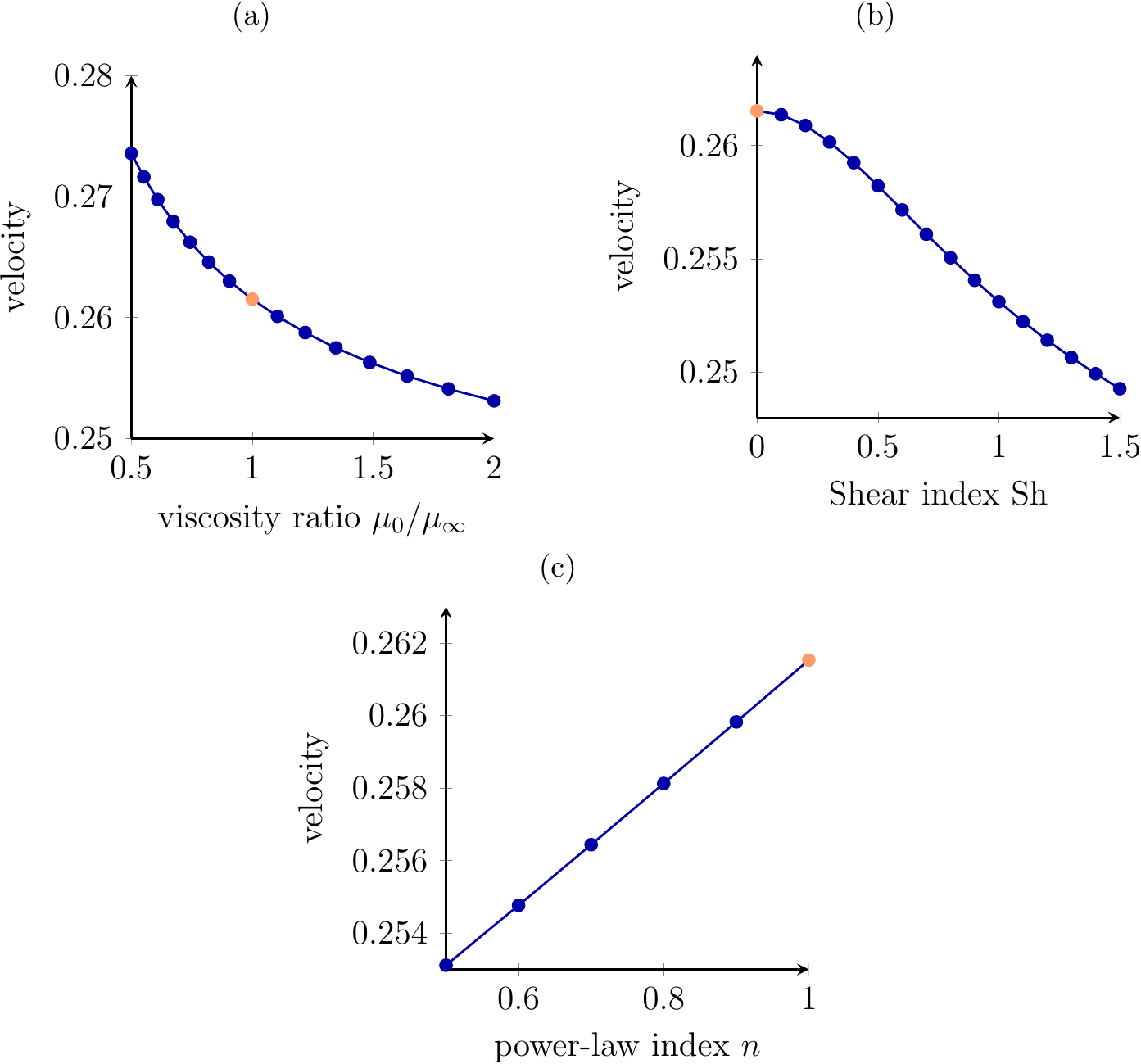}
\caption{The velocity of the treadmilling squirmer with slip velocity given by
equation \eqref{eq:squirmer_param} as a function of (a) the viscosity ratio
$\mu_0/\mu_\infty$ with $n = 0.5$ and $\mathrm{Sh} = 1$, (b) the shear index
$\mathrm{Sh}$ with $n = 0.5$ and $\mu_0/\mu_\infty = 2$ and (c) the power-law
index $n$ with $\mu_0/\mu_\infty = 2$ and $\mathrm{Sh} = 1$. In each panel, the
case corresponding to Newtonian fluid is marked in lighter gray (orange online).} 
\label{fig:squirmer_treadmill} 
\end{center}
\end{figure}

Shear-thinning decreases the velocity of this squirmer (figure
\ref{fig:squirmer_treadmill}). This result draws an interesting
parallel with the work of \citet{zhu2012self}, who found that spherical
squirmers were also hindered by a different non-Newtonian fluid property,
viscoelasticity.  Figure \ref{fig:squirmer_treadmill}c shows a
striking apparently linear dependence of the swimming velocity upon the
power-law index $n$. The decrease in  velocity is small; for $\mu_0/\mu_\infty =
2, n = 0.5$ and $\mathrm{Sh} = 1$, the velocity is reduced by a little over
$3\%$.

\begin{figure}[tbp]
\begin{center}
\includegraphics[scale = 0.83333]{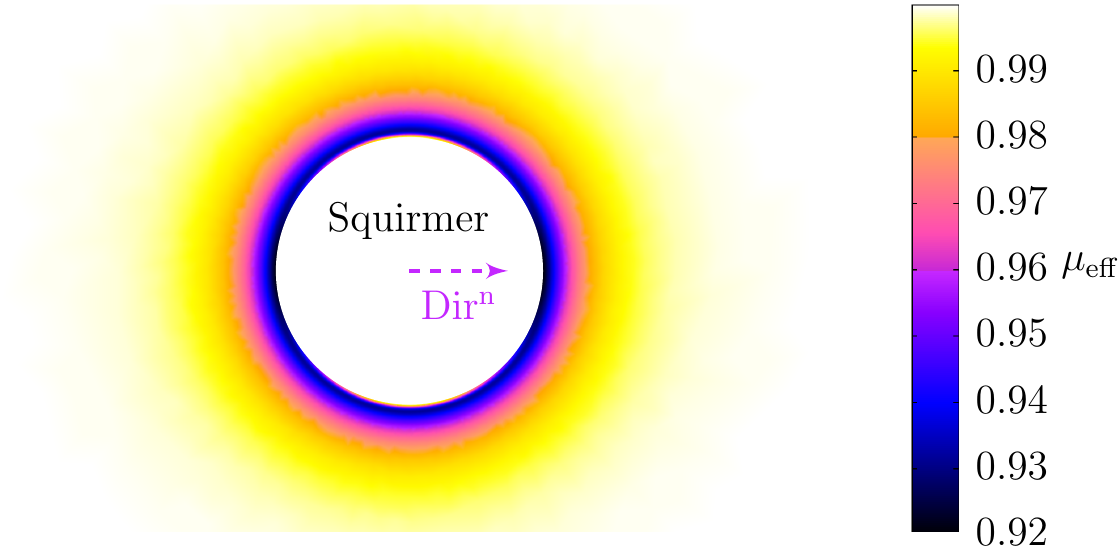}
\caption{The effective viscosity {$\mu_\mathrm{eff}$} of Carreau fluid,
normalized to $\mu_0 = 1$, surrounding the treadmilling squirmer for
$\mu_0/\mu_\infty = 2, n = 0.5$ and $\mathrm{Sh} = 0.5$. These parameter values
are the extremal values used for the data in figures
\ref{fig:squirmer_viscosity_radial} and \ref{fig:envelope_visc_vel_squirmer}.
Away from the swimmer surface, contours of equi-viscosity are approximately
circular. On the surface, fluid is relatively thicker surrounding the propulsive
portions of the swimmer. The squirmer is aligned to the positive $x$-axis, as
in figure \ref{fig:squirmer_volvox}a, and the direction of travel is
indicated by the dashed arrow.}
\label{fig:squirmer_viscosity}
\end{center}
\end{figure}

The effective viscosity field of the flow has a simple form; even relatively
near to the swimmer, contours of equi-viscosity are approximately circular,
centered on the swimmer (figure \ref{fig:squirmer_viscosity}).  However, very
near to the surface, the fluid surrounding the propulsive elements of the
treadmilling squirmer is relatively thicker than that surrounding the
drag-inducing portions.  Thus, the viscosity differential for this squirmer is
positive, yet its velocity is decreased by shear-thinning, demonstrating that
slip velocity models differ from no-slip multiple sphere swimmers in this
respect. The reduction in velocity arises from the envelope of thinned fluid
surrounding the squirmer.

\begin{figure}[tbp]
\begin{center}
\includegraphics[scale = 0.83333]{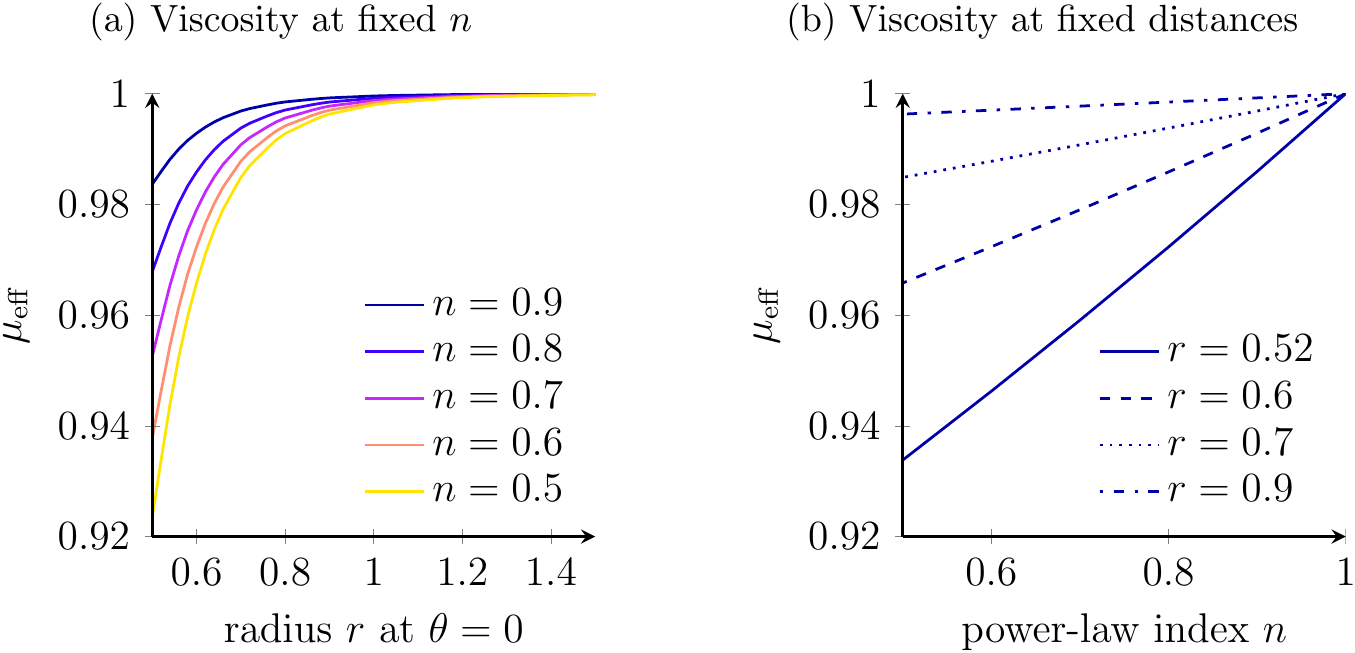}
\caption{The effective viscosity of the fluid envelope surrounding the
treadmilling squirmer. (a) Changes in the viscosity field as a function of the
radial coordinate $r$ for different values of the power-law index $n$. The
swimmer surface is given by $r = 0.5$.  (b) For fixed values of $r$, the
effective viscosity exhibits a near linear dependence upon the power-law index
$n$.}
\label{fig:squirmer_viscosity_radial}
\end{center}
\end{figure}

Figure \ref{fig:squirmer_viscosity_radial} shows the radial variation in the
effective viscosity of the fluid surrounding the squirmer. As $n$ decreases, the
viscosity immediately surrounding the swimmer decreases, but the rate at which
the viscosity approaches the zero-shear value increases. As a result of this
increase, the size of the envelope of thinned fluid surrounding the swimmer
varies little with changes in rheological parameters (figure
\ref{fig:squirmer_viscosity_radial}a). For any fixed value of the radial
coordinate $r$, with $r = 0.5$ being the squirmer's surface, the effective
viscosity at that point decreases approximately linearly with $n$ (figure
\ref{fig:squirmer_viscosity_radial}b). 

\begin{figure}[tbp]
\begin{center}
\includegraphics[scale = 0.83333]{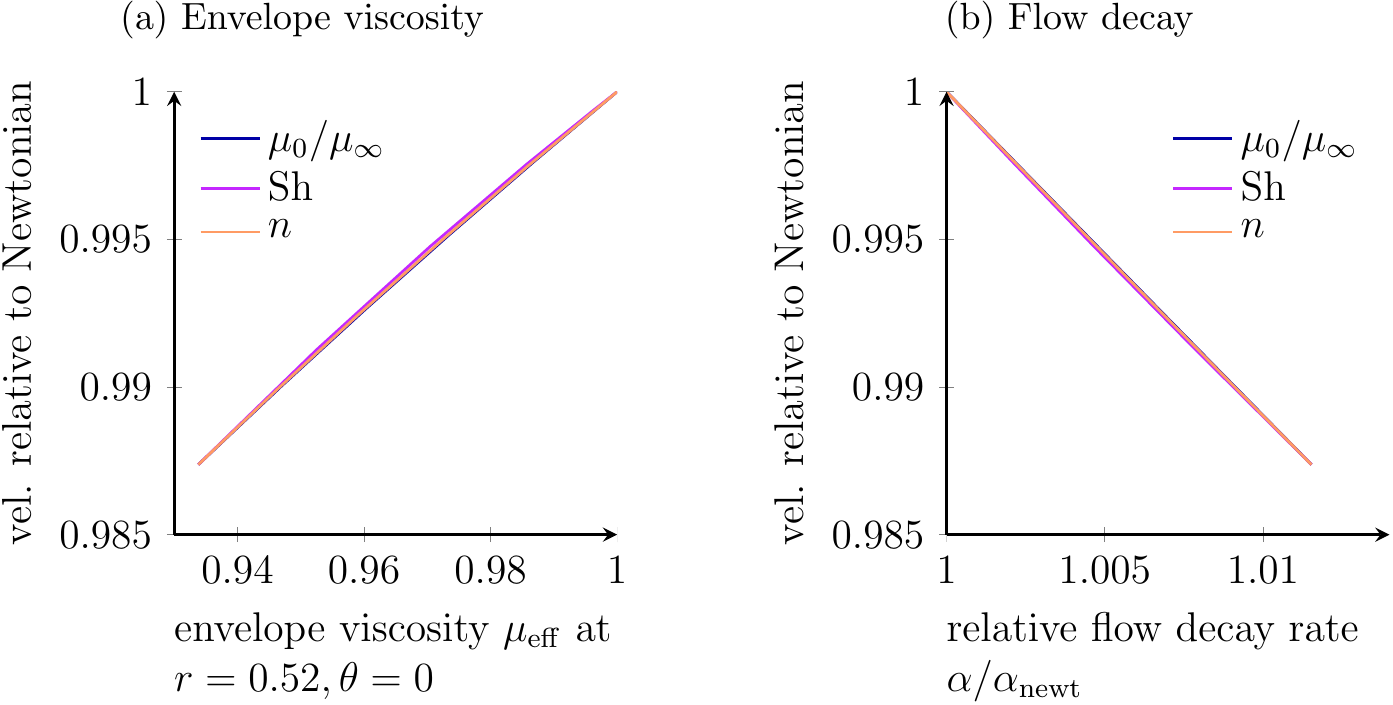}
\caption{The velocity relative to the Newtonian case of the treadmilling
squirmer as a function of (a) the effective viscosity on the contour $r = 0.52$
and (b) the rate of decay $\alpha$ of the velocity from the surface of the
squirmer relative to the Newtonian case $\alpha_{\mathrm{newt}}$. The velocity
has been calculated while varying the three rheological parameters of Carreau
flow for $n =0.5, \mu_0/\mu_\infty \in [1, 2], \mathrm{Sh} = 0.5$ (dark gray,
blue online), $n = 0.5, \mu_0/\mu_\infty = 2, \mathrm{Sh} \in [0, 0.5]$ (medium
gray, magenta online) and $n \in [0.5, 1], \mu_0/\mu_\infty = 2, \mathrm{Sh} =
0.5$ (light gray, orange online). This figure demonstrates a striking
proportionality between the velocity and the decay rate of the fluid.}
\label{fig:envelope_visc_vel_squirmer}
\end{center}
\end{figure}

Since the decrease in swimming velocity also exhibits a linear dependence upon
the power-law index $n$, we examine the dependence of swimming velocity on the
effective viscosity of the fluid surrounding the squirmer. Figure
\ref{fig:envelope_visc_vel_squirmer}a shows the decrease in swimming velocity
relative to the Newtonian case as a function of the effective viscosity of the
fluid envelope at $r = 0.52$, a small distance from the squirmer's surface, for
varying viscosity ratio, shear index and power-law index. This figure
demonstrates a strong linear correlation between the effective viscosity of the
fluid a small distance from the swimmer's surface and the swimmer's velocity.

However, whilst the absolute values of viscosity do not affect swimmers with
prescribed kinematics, the envelope of thinned fluid shields the far field flow
from the flow generated by the squirmer. As fluid becomes relatively thinner
around the squirmer, the decay rate of the near-field flow increases. This draws
an interesting parallel with the work of \citet{zhu2012self}, who found a
similar effect for viscoelastic (Giesekus) fluids. In the near-field, along the
line $\theta = 0$, the velocity of the flow is approximately
\begin{equation}
u \approx \frac{A}{r^\alpha},\quad \therefore \log u \approx \log A - \alpha \log r.
\end{equation}
Thus, the flow decay rate is given by 
\begin{equation}
\alpha = - \frac{\Delta \log u}{\Delta \log r}.
\end{equation}
Close to the squirmer's surface, the Newtonian flow decay rate
$\alpha_{\mathrm{newt}} = 1.95$.

Figure \ref{fig:envelope_visc_vel_squirmer}b shows the swimming velocity of the
squirmer as a function of this decay rate at $r = 0.52, \theta = 0$, a small
distance from the squirmer's surface, relative to the Newtonian case for varying
rheological parameters $\mu_0/\mu_\infty, n$ and $\mathrm{Sh}$. The decrease in
velocity and increase in flow decay exhibit a linear relationship, and are the
same magnitude; the slope of the curve is close to $-1$. This observation
motivates the following argument: The squirmer generates an envelope of thinned
fluid around itself when swimming through Carreau fluid. This envelope increases
the decay rate of flow away from the squirmer's surface.  Thus, prescribed
motion on the surface moves relatively less fluid, which decreases the swimming
velocity.

However, models of squirmers exhibiting surface velocity distribution may
neglect effects arising from rheological interactions at the scale of individual
cilia. These interactions may be captured more effectively by squirming models
for which the surface is subject to small deformations. For many ciliates, such
as the protozoa \textit{Opalina}, surface deformation provides a better
representation of the swimmer than slip velocity modeling.  It may be that
rheologically-enhanced propulsion at the cilium scale is captured by envelope
models with surface deformation. 


\subsection{Monoflagellate pushers}
\label{subsec:flagellate_swimmers}

\begin{figure}[tbp]
\begin{center}
\includegraphics[scale = 0.83333]{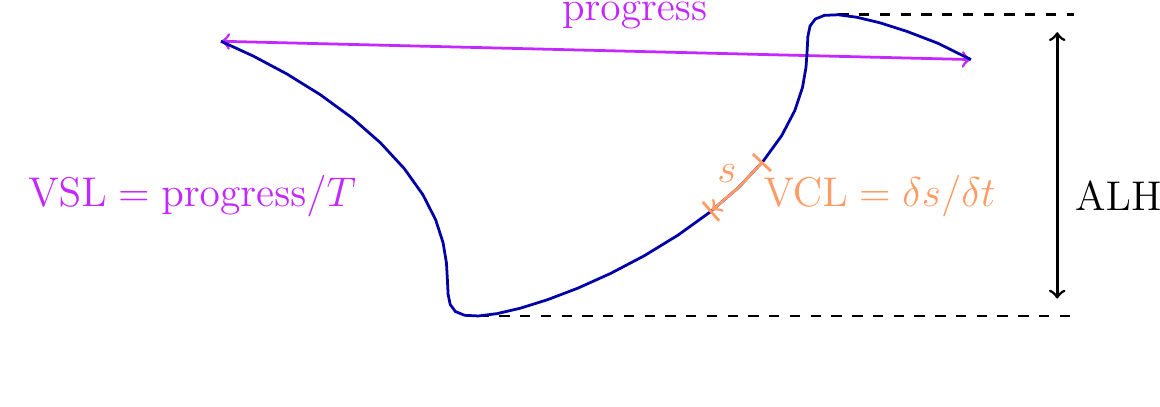}
\caption{Swimming parameters for the trajectory (dark gray, blue online) of a
swimmer moving from right to left over one beat cycle of period $T$. The
instantaneous velocity is the derivative of arclength $s$ along the path with
respect to time.}
\label{fig:sperm_swimming_parameters}
\end{center}
\end{figure}

We will now examine the effects of shear-thinning rheology on the swimming of a
two-dimensional model sperm with prescribed waveform. Since the trajectories of
such swimmers are two-dimensional, we will analyze their shape using variables
from Computer Aided Semen Analysis (CASA), see for example
\citet{mortimer1997critical}. Our usage will differ slightly, in that CASA
variables are statistical averages over many beat cycles determined from video
microscopy of living cells sampled at a given frequency, whereas we will
generate a smooth, time periodic waveform and thus our parameters will be
measured over a single beat. The variables we will consider are demonstrated
for an example trajectory over one beat cycle in figure
\ref{fig:sperm_swimming_parameters}. 

Sperm do not exhibit an effective-recovery stroke pattern, but rather swim by
propagating a travelling wave along the flagellum. As such, we now refer to a
swimmer's `progress' as the distance between its start and end points over a
beat. We will also consider its straight line velocity $\mathrm{VSL} =
\mathrm{progress}/T$ and its curvilinear, or instantaneous, velocity
$\mathrm{VCL}$, the velocity of the cell at any given point in time. The
amplitude of the cell's lateral head displacement $\mathrm{ALH}$, is given by
the difference between the maximum and minimum $y$ values on the trajectory. We
also consider the path length $\mathrm{PL}$ of the trajectory, that is the total
distance travelled, as well as the straightness of the path $\mathrm{STR} =
\mathrm{progress}/\mathrm{PL}$.

\begin{figure}[tbp]
\begin{center}
\includegraphics[scale = 0.83333]{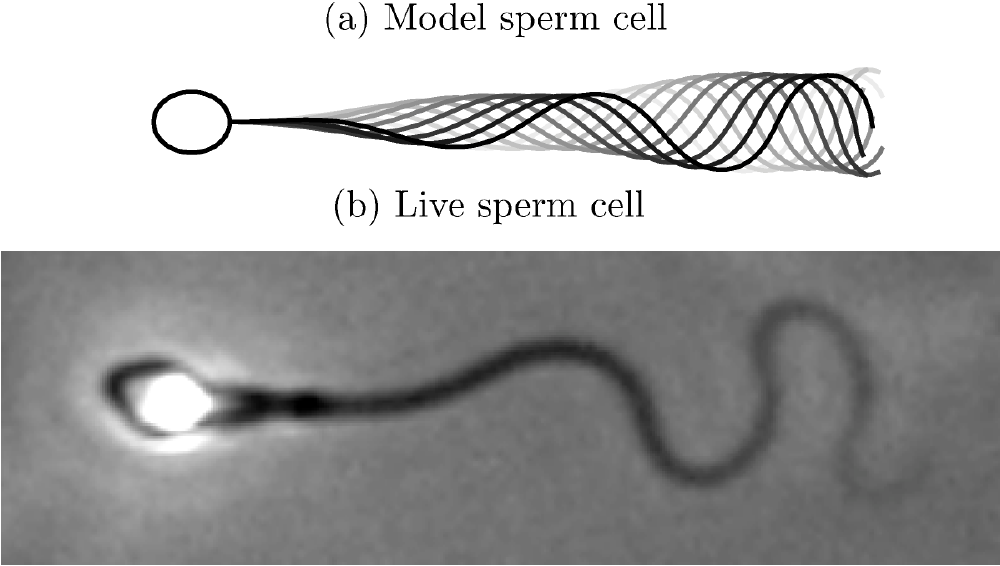}
\caption{(a) The flagellar waveform generated by shear angle
\eqref{eq:shear_wave} and (b) a micrograph of a human sperm in medium containing
$1\%$ methylcellulose, a fluid with comparable viscosity to that of cervical
mucus.}
\label{fig:sperm_shear_waveform} 
\end{center}
\end{figure}

The swimmer is propelled by a single flagellum that propagates a bending wave
along its length, generating the forces required to move the cell forward. We
parameterize the flagellum in terms of its shear angle $\psi(s,t)$ given in the
body frame. A shear angle of the form
\begin{equation}
	\psi (s,t) = C s \cos[2\pi(ks - \omega t)], \label{eq:shear_wave}
\end{equation}
represents a bending wave propagating down the flagellum, steepening towards the
less stiff distal end with a linear envelope. This shear angle produces a
waveform representative of sperm swimming in high viscosity fluids
\citep{smith2009bend}, shown in figure \ref{fig:sperm_shear_waveform}. The lab
frame position of the flagellum is then given by rotating the centerline in the
body frame by the swimmer's orientation, and translating by the current head
position.

Length scales are normalized to the flagellum length, so that one length unit
corresponds to $55\,\mu\mathrm{m}$, and one time unit corresponds to a single
beat of the flagellum. Thus, for a tail beating at $10\,\mathrm{Hz}$ one time
unit corresponds to $0.1\,\mathrm{s}$.

\citet{johnson2012modelling} showed that particular sperm-like swimmer
progressed further in shear-thinning fluids. In this study, we will show that
this behavior arises for other sperm-like swimmers, and examine the interplay
between physical mechanisms and morphological changes in swimming trajectory
that cause it.

We will examine the trajectories of swimmers with waveforms generated by the
shear angle \eqref{eq:shear_wave} for maximum shear angle $A = 0.45\pi$ and
wavenumber $k = 2.5$, i.e. $2.5$ waves on the flagellum. We have also examined
waveforms produced by other parameter values, and found that the effects of
shear-thinning were consistent for all values considered. The cell head will be
given by an ellipse of fixed eccentricity, but different area, given in table
\ref{tab:head_sizes}.
 
\begin{table}
	\begin{center}
	\begin{tabular}{@{}llcr@{}} \toprule
		\multicolumn{4}{c}{Sperm head morphologies} \\
		$a_x$	& $a_y$  & Area & Circumference \\ \midrule
		0.045	& 0.036 & 0.0016$\pi$  & 0.255 \\
		0.05	& 0.04 & 0.002$\pi$ & 0.284 \\
		0.055 & 0.044 & 0.0024$\pi$ & 0.312\\\bottomrule
	\end{tabular}	
	\end{center}
	\caption{Elliptical head morphologies of constant eccentricity, but
	different area scaled with flagellum length, corresponding to the data in figure
	\ref{fig:sperm_head}. These morphologies, from top to bottom correspond
	with dark to light plots.}
	\label{tab:head_sizes}
\end{table}

\begin{figure}[tbp]
\begin{center}
\includegraphics[scale = 0.83333]{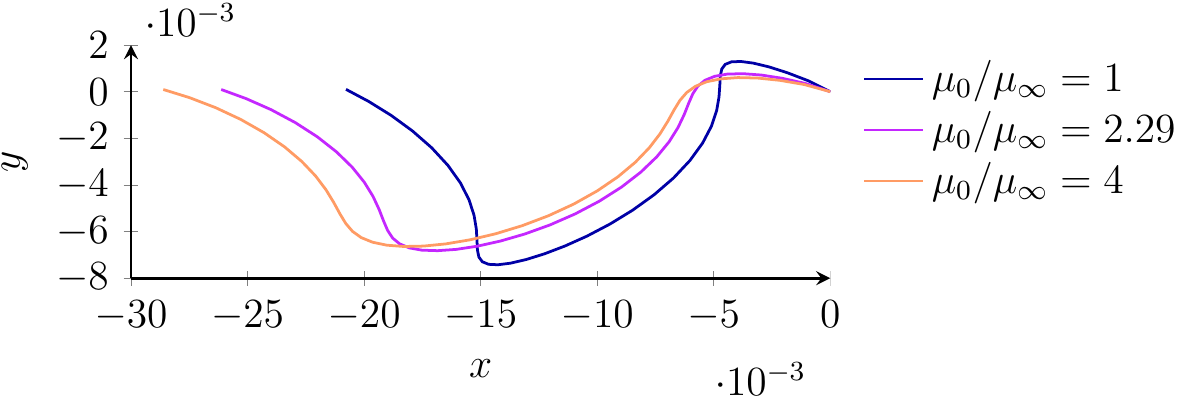}
\caption{Trajectories of the body frame origin $\mathbf{x}_0$, given by the
head-flagellum junction, of a two-dimensional sperm-like swimmer in Carreau fluid
for different values of the viscosity ratio $\mu_0/\mu_\infty$, showing an
increase in progress and a decrease in $\mathrm{ALH}$ as $\mu_0/\mu_\infty$
increases.} 
\label{fig:sperm_trajectories}
\end{center}
\end{figure}

Figure \ref{fig:sperm_trajectories} shows the trajectories of an example sperm
for three values of the viscosity ratio. From this figure, it is apparent that
shear-thinning increases the progress of sperm-like swimmers significantly;
for $\mu_0/\mu_\infty = 4, n = 0.5$ and $\mathrm{Sh} = 1$, this increase is
around $40\%$ over the Newtonian case. However, it is not immediately apparent
how much of the increase in progress is associated with increased path
straightness ($\mathrm{STR}$) and how much arises from increased instantaneous
velocity ($\mathrm{VCL}$).

\begin{figure}[tbp]
\begin{center}
\includegraphics[scale = 0.83333]{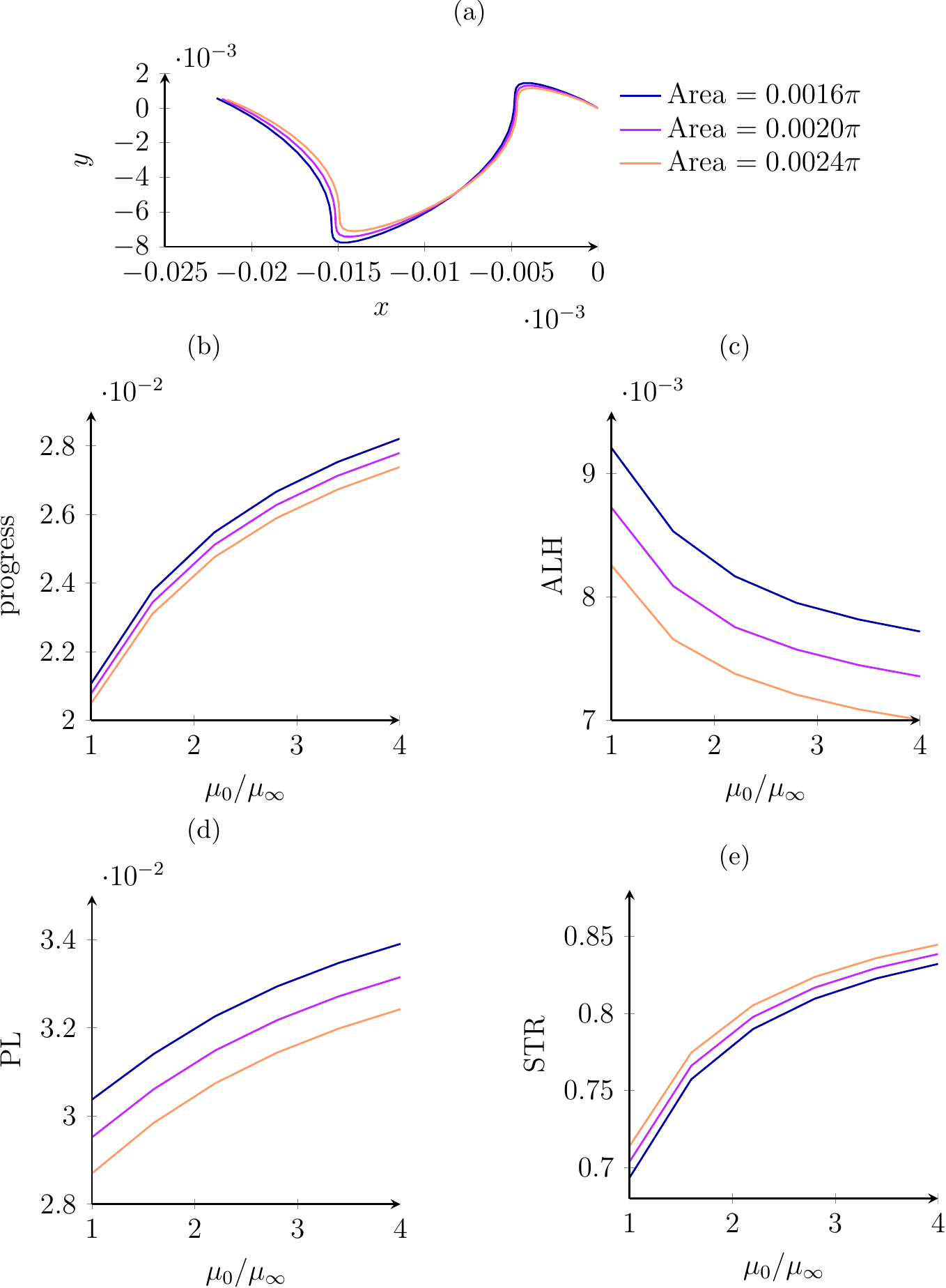}
\caption{(a) Trajectories of the cells with head morphologies given in table
\ref{tab:head_sizes}, swimming in Stokes flow with $n = 0.5$, $\mu_0/\mu_\infty
= 4$ and $\mathrm{Sh} = 1$. For $n = 0.5$ and $\mathrm{Sh} = 1$, the effect of
varying the viscosity ratio $\mu_0/\mu_\infty$ on (b) the swimmers' progress,
(c) the amplitude of the swimmers' lateral head displacement, (d) the path
length of the swimmers' trajectories and (e) the swimmers' path straightness.}
\label{fig:sperm_head} 
\end{center} 
\end{figure}

Figure \ref{fig:sperm_head} demonstrates the effects of shear-thinning on the
shape of the swimming trajectory for sperm with the three different head sizes
given in table \ref{tab:head_sizes}. The trajectories that these swimmers with
different head sizes follow in Stokes flow are shown in figure
\ref{fig:sperm_head}a, showing that increasing head size leads to a small
decrease in progress, due to increased drag. Shear-thinning increases progress
(figure \ref{fig:sperm_head}b) by reducing the side-to-side motion of the cell
$\mathrm{ALH}$ (figure \ref{fig:sperm_head}c) but increasing its instantaneous
velocity $\mathrm{VCL}$, as reflected by increased path length $\mathrm{PL}$
(figure \ref{fig:sperm_head}d). This increases the swimmer's path straightness,
$\mathrm{STR}$, shown in figure \ref{fig:sperm_head}e, which is apparent when
the trajectories of a single swimmer, with $a_x = 0.05$ and $a_y = 0.04$, for
various values of the viscosity ratio are plotted together (figure
\ref{fig:sperm_trajectories}).

These effects are robust to morphological and kinematic changes. Varying the
eccentricity of the cell head or the wavenumber changes the swimmer's
trajectory, but the rheological effects that we show are consistent with changes
in these parameters. To understand the increase in cell progress, we will now
examine the viscosity field surrounding the swimmer, and the force generated by
the flagellum.

\begin{figure}[tbp]
\begin{center}
\includegraphics[scale = 0.83333]{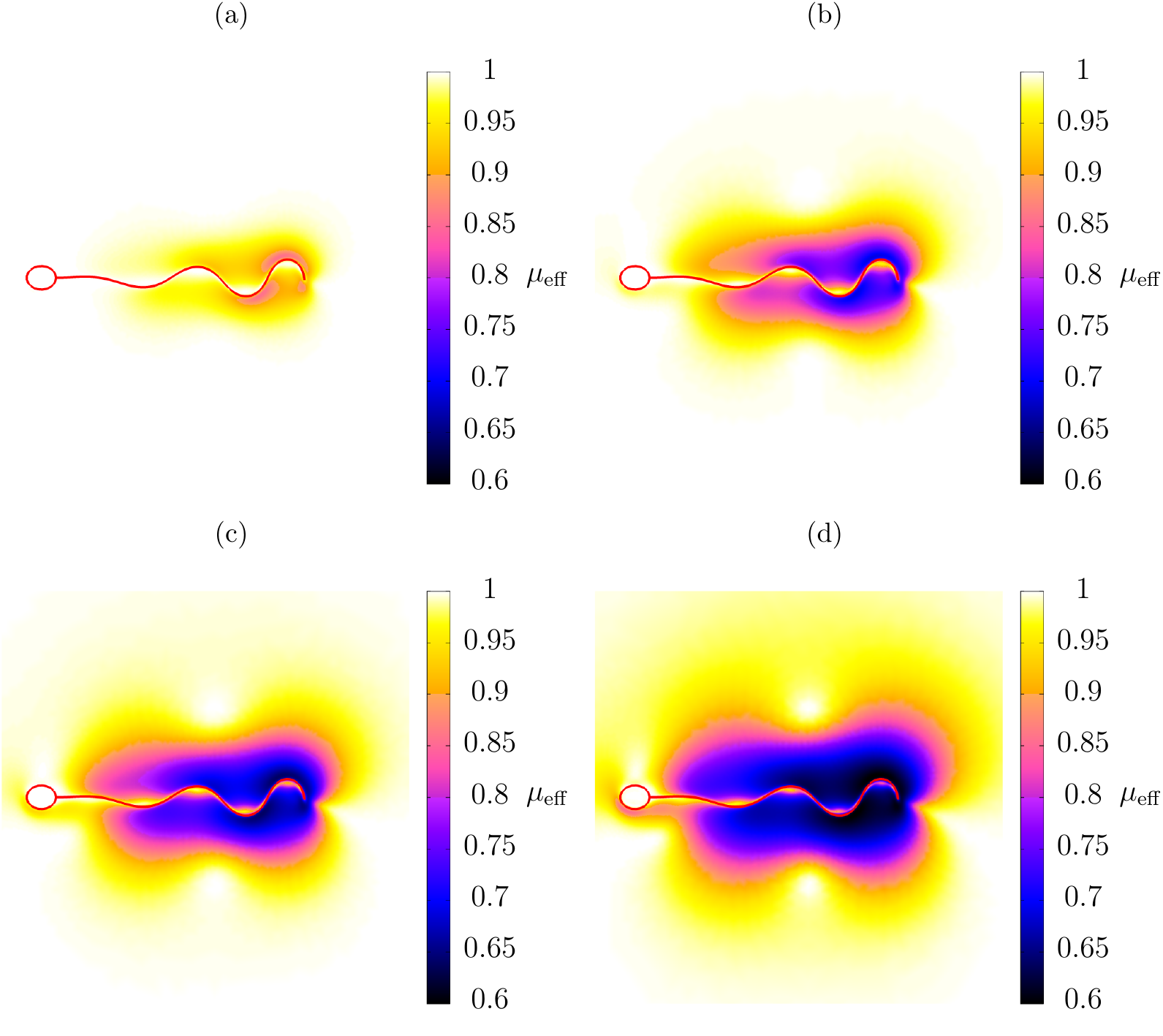}
\caption{The impact of varying $\mathrm{Sh} = \lambda\omega$ on the effective
viscosity $\mu_{\mathrm{eff}}$ of Carreau fluid surrounding a two-dimensional
sperm-like swimmer at (a) $\mathrm{Sh} = 0.2$, (b) $\mathrm{Sh} = 0.8$, (c)
$\mathrm{Sh} = 1.5$ and (d) $\mathrm{Sh} = 3$ with $\mu_0/\mu_\infty = 2$ and $n
= 0.5$. In these figures, the area of the cell head is $0.002\pi$, the
wavenumber $k = 2.5$ and the maximum shear angle $A = 0.45\pi$.}		
\label{fig:sperm_visc_field}
\end{center}
\end{figure}

The viscosity field surrounding the swimmer is shown for four values of the
shear index $\mathrm{Sh}$ in figure \ref{fig:sperm_visc_field}. Fluid is
thickest around the cell head, and there is a gradient of thick to thin fluid
along the flagellum, as well as the slightly less obvious feature of a gradient
of thick to thin fluid across the swimmer which alternates in sign at local
maxima of the shear angle $\psi$. As $\mathrm{Sh}$ is increased to an optimum
value these gradients are enhanced, after which they decrease because the fluid
becomes thinned substantially at the head end of the flagellum.
\citet{johnson2012modelling} found an optimal value of $\mathrm{Sh}$ for a
particular sperm-like swimmer's progress. We now find that this optimal progress
is associated with maximal gradients along the flagellum.

\begin{figure}[tbp]
\begin{center}
\includegraphics[scale = 0.83333]{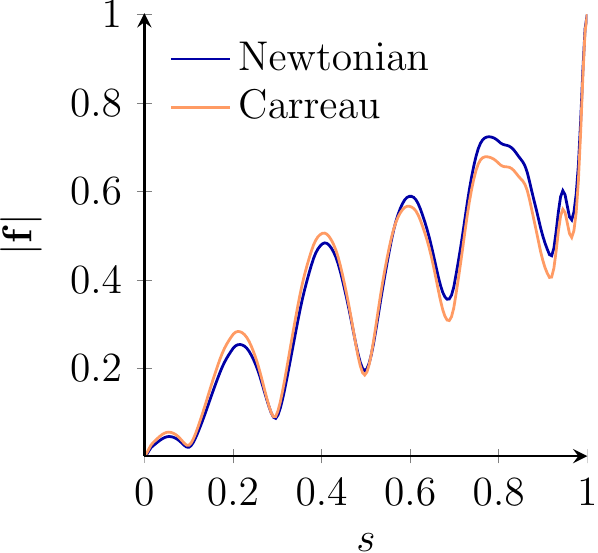}
\caption{The magnitude of the force that the flagellum exerts upon the fluid at
time $t = 0$ for Newtonian (dark gray, blue online) and Carreau (light gray,
orange online) fluids with $\mu_0/\mu_{\infty} = 2, n = 0.5$ and $\mathrm{Sh} =
0.8$, close to the optimal value of $\mathrm{Sh}$ found by
\citet{johnson2012modelling}.} 
\label{fig:sperm_force_magnitude}
\end{center}
\end{figure}

\begin{figure}[tbp]
\begin{center}
\includegraphics[scale = 0.83333]{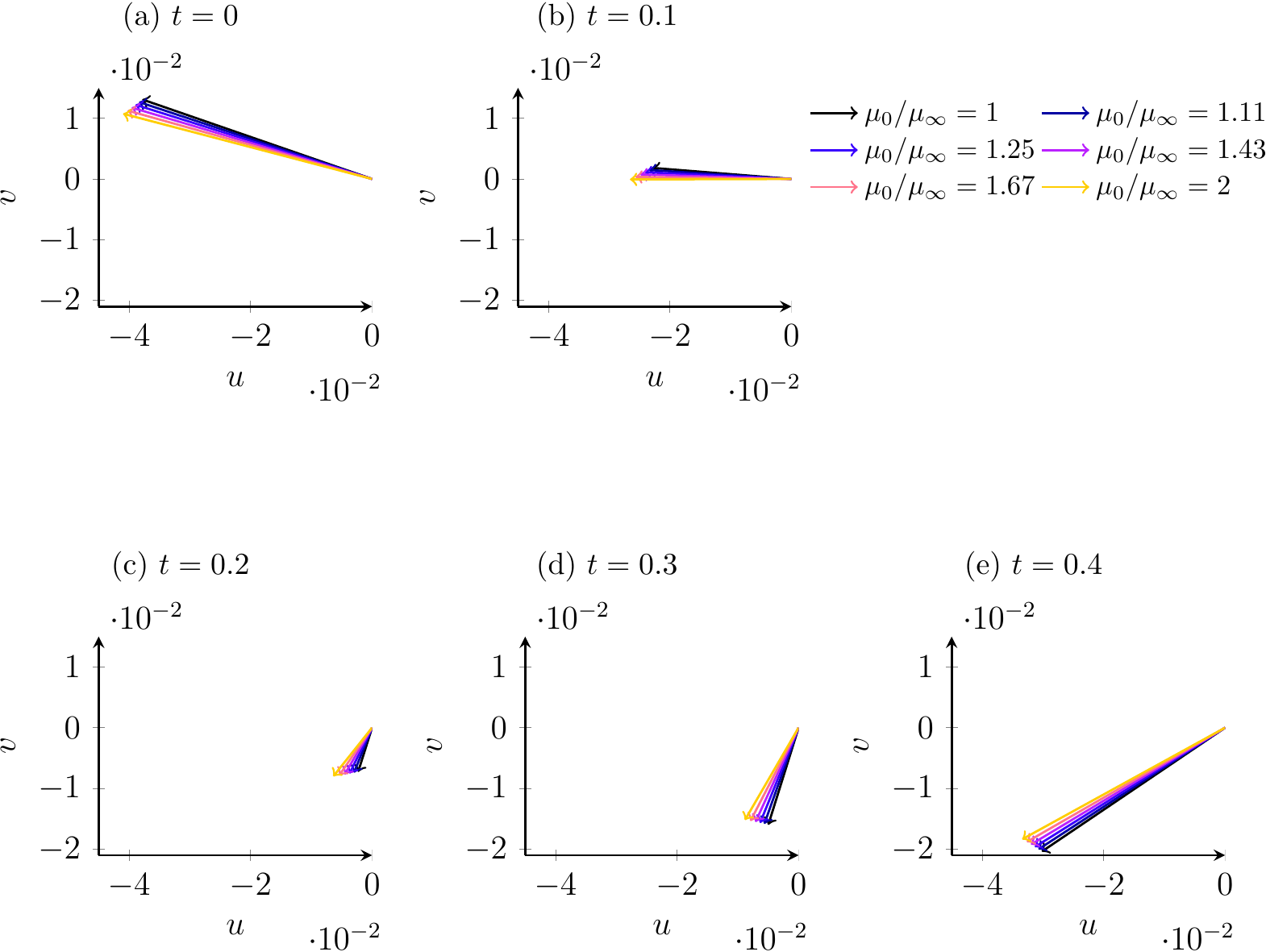}
\caption{The magnitude and direction of swimming of a sperm oriented in the
negative $x$ direction with wavenumber $k = 2.5$ and maximum shear angle $A =
0.45\pi$ at times $t = 0, 0.1, 0.2, 0.3, 0.4$, for varying viscosity ratio.
These times span half a complete beat cycle. This figure demonstrates that
shear-thinning results in straighter swimming and increased instantaneous
velocity.} 
\label{fig:sperm_swimming_magnitude_direction}
\end{center}
\end{figure}

We examine the forces exerted by the flagellum on the fluid at five equally
spaced instants over half its beat cycle for varying viscosity ratio. At each
moment, the gradient of thick to thin fluid along the flagellum that arises in
shear-thinning fluids entails that forces generated in the proximal (near to
head) portion of the flagellum have greater magnitude relative to those in the
distal (near to tip) portion, when compared to the Newtonian case (figure
\ref{fig:sperm_force_magnitude}). Thus, shear-thinning induces a redistribution
of force from the distal to the proximal end of the flagellum. This
redistribution has the effect of making the force distribution more symmetric
about the body axis, and thus straightens the trajectory. This effect is shown
in figure \ref{fig:sperm_swimming_magnitude_direction}, where the magnitude and
direction of swimming have been plotted for a sperm aligned with the negative
$x-$axis at times $t = 0, 0.1, \dots 0.4$ for changing values of the viscosity
ratio. Figure \ref{fig:sperm_swimming_magnitude_direction} also demonstrates the
increase in the magnitude of instantaneous velocity resulting from
shear-thinning rheology. The increased instantaneous velocity acts in concert
with the straightened path to yield significant increases in progress. 

\section{Discussion}
\label{sec:discussion}

We have analyzed the effects of shear-thinning rheology on three distinct
classes of microscopic swimmer with prescribed kinematics in Carreau fluid. This
continuum approach to modeling biological fluids may not be appropriate when the
swimmer and the suspended fibers are of a comparable length, as with bacteria in
mucus \citep{lai2009micro}, but it can still provide insight into important
effects.

Whilst our modeling is two-dimensional, the observed physical effects are likely
to be present for three-dimensional swimmers: sliding spheres exert stresses on
the fluid, thereby thinning a surrounding envelope. The Najafi-Golestanian
swimmer payload travels more slowly through the fluid than its propulsive
sphere. Consequently, fluid surrounding the propulsive sphere is thinned more
than fluid surrounding the payload spheres, resulting in a decrease in
instantaneous velocity. By contrast, the paddler payload moves more quickly
through the fluid than the propulsive elements, fluid around the payload is
relatively thinner, thereby increasing instantaneous velocity. The relatively
higher decay rate of three-dimensional flow will be associated with an increased
decay of the viscosity field around each sphere. This decay will in turn reduce
the asymmetry between the effects of shear-thinning on the effective and
recovery strokes. So while shear-thinning will decrease both the progress and
regress of a Najafi-Golestanian swimmer, we therefore predict that the increase
in net progress will be relatively less than for an equivalent two-dimensional
swimmer.

The squirmer in three dimensions will again thin an envelope of surrounding
fluid, enhancing flow decay rate and thereby decreasing swimming velocity in
shear-thinning fluids. For sperm-like swimmers, the prescribed waveforms we
considered increase in velocity to the distal portion of the flagellum, and are
therefore likely to generate a gradient of thick to thin fluid along the
flagellum as in the two-dimensional case. However, in three dimensions, fluid
can also pass over the flagellum, and so this gradient may be reduced.

The effects found also give insight into sliding sphere swimmers that may
violate Purcell's Scallop theorem. Since the instantaneous velocity of the
sliding sphere swimmers analyzed is approximately proportional to the viscosity
differential, an asymmetry between the body frame speed of effective and
recovery strokes should allow a reciprocal swimmer to progress through
inertialess Carreau fluid. Net progress is made possible because faster motion
thins the fluid to a greater extent, thereby inducing an asymmetry between the
effective and recovery flow viscosity fields. In Newtonian fluid, no such
asymmetry arises, and due to the time independence in the governing equations,
such reciprocal motion will not result in net progress.

\begin{figure}[tbp]
\begin{center}
\includegraphics[scale = 0.83333]{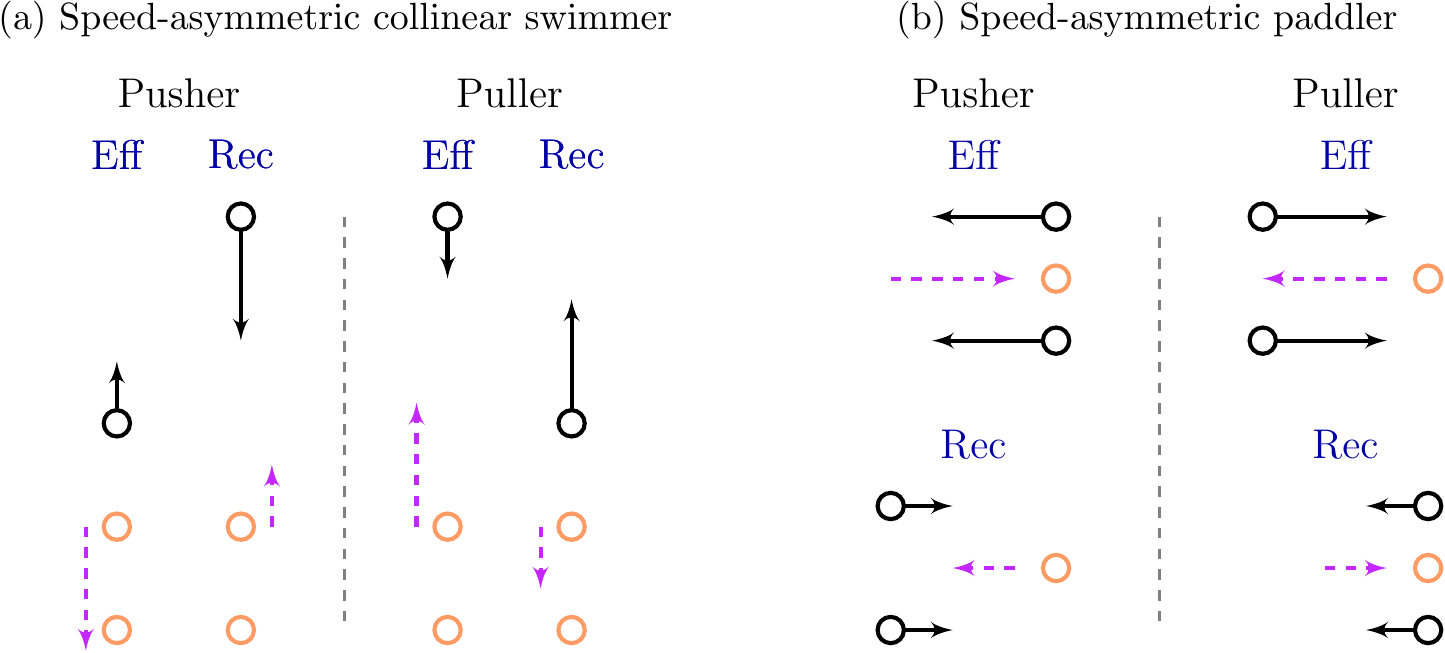}
\caption{Reciprocal sliding sphere swimmers that cannot progress through
inertialess Newtonian fluid, but may progress through inertialess Carreau fluid.
These swimmers are pusher and puller versions of (a) the Najafi-Golestanian
swimmer and (b) the paddler, showing the effective and recovery strokes with an
indication of the velocity of the propulsive sphere (solid arrow) and the
magnitude and direction of progress over each stroke (dashed arrow).}
\label{fig:reciprocal_sliding_sphere}
\end{center}
\end{figure}

Two such reciprocal swimmers may be formed from each of the Najafi-Golestanian
swimmer and the three-sphere paddler, as shown in figure
\ref{fig:reciprocal_sliding_sphere}. We refer to these models as the
speed-asymmetric collinear swimmer and paddler respectively.  For each
speed-asymmetric swimmer, a ``pusher'' and ``puller'' version of the swimmer may
be modeled: pushers are swimmers whose payload is pushed from behind, such as
most animal sperm, whereas pullers, such as algae are pulled from the front.

The net propulsion due to stroke speed asymmetry is, however, very slight. For
the speed-asymmetric collinear pusher described in table
\ref{tab:pusher_golestanian}, simulations in a channel of length $20L$ were
performed to minimize boundary truncation effects, and for fixed $\mathrm{Sh} =
1, \mu_0/\mu_\infty = 2$, net progress over a beat was maximized at $0.001L$ for
$n = 0.7$, which is approximately $0.2\%$ of the body frame beat amplitude. This
is in contrast to the Najafi-Golestanian swimmer given in table
\ref{tab:three_sphere}, which progresses approximately $10\%$ of its amplitude
per beat. The difference between pushers and pullers was not discernible to
within the resolution of our method.

\begin{table}
\begin{center}
\begin{tabular}{@{}lcccr@{}} \toprule
\multicolumn{5}{c}{Speed-asymmetric collinear pusher} \\
Stroke  & $x_1$			& $x_2$ & $x_3$ & time $t$ \\ \midrule
Eff 	& $-(d-a) - 8at/3$     	& $0$   & $d-a$ & $[0,3/4)$  \\         
Rec 	& $-(d+a) + 8a(t-3/4)$       	& $0$   & $d-a$ & $[3/4,1)$ \\ \bottomrule
\end{tabular}
\end{center}
\caption{The body frame positions of the three spheres of the
speed-asymmetric collinear pusher over its effective stroke, which lasts for
$3/4$ of the beat period, and the recovery stroke, which lasts for $1/4$ of the
beat period.}
\label{tab:pusher_golestanian}
\end{table}

Instead of a kinematic description, sliding sphere swimmers may also be defined
in terms of a prescribed force. Whilst we will not fully examine this question
in this work, it is interesting to consider how shear-thinning would affect such
a swimmer. The above reasoning and methodology can be used to provide insight
into these effects. For example, during the effective stroke of a
Najafi-Golestanian swimmer, the swimming arm exerts a prescribed force on the
fluid which is independent of viscosity. By force balance, this propulsive force
is equal to the drag force on the payload.  However, in shear-thinning fluid,
the payload thins an envelope of surrounding fluid, which decreases is drag
coefficient, thereby increasing the swimming speed for a given drag force. Thus,
our results suggest that the instantaneous velocity of a prescribed force
Najafi-Golestanian swimmer may increase with shear-thinning: the opposite
behavior to that of the prescribed kinematic swimmer. More complex regulation of
swimmer beating will be an interesting avenue of future research.


\section{Conclusions}
\label{sec:conclusions}

Shear-thinning is an important property of many biological fluids. In this
paper, we found that its effects upon microscopic swimmers are highly sensitive
to the swimming stroke employed. The collinear sliding sphere swimmer
experiences decreases in instantaneous velocity during both effective and
recovery strokes, but increases in net progress; the opposite effect occurs for
the paddler. A slip-velocity squirmer was hindered by shear-thinning, and
sperm-like swimmers were aided by it. The magnitudes of these effects were small
(of order $3 \%$) for sliding sphere swimmers and squirmers, but could be larger
(of order $10 \%$) for sperm-like swimmers.

The effects of shear-thinning on sliding sphere swimmers can be understood by
considering the viscosity differential, provided the spheres are sufficiently
separated. Positive viscosity differential entails thicker fluid around the
propulsive spheres relative to the payload, increasing instantaneous velocity
and vice-versa. When spheres are closer together, the envelope of thinned fluid
surrounding the swimmer hinders swimming, as with the squirmer. This envelope
resulted in a smaller increase in velocity during the effective
stroke than during the recovery stroke of the paddler, reducing net progress.
The same effect induced a greater decrease in velocity during the recovery
stroke of the Najafi-Golestanian swimmer, increasing net progress.

The envelope of thinned fluid surrounding the squirmer was shown to reduce the
swimmer's instantaneous velocity. This reduction was associated with enhanced
flow decay within the thinned envelope. However, the envelope approach of
time-averaging the coordinated action of many cilia into a surface slip velocity
might neglect rheological interactions that occur on the scale of each cilium,
and thus it may be desirable in the future to consider squirming models
exhibiting small surface deformations, or models incorporating discrete cilia.

Sperm-like swimmers induced a gradient of thick to thin fluid along their
flagellum, which was associated with both a flattening of the swimming
trajectory and an increase in instantaneous velocity. These effects were
complementary, leading to significant increases in progress per beat.

Finally, we suggested two model reciprocal swimmers comprising sliding spheres
which achieve progression through Carreau fluid by manipulating the viscosity
differential. This effect results from speed asymmetry between the effective and
recovery strokes. However, the net progress achieved over a beat is slight; the
net progress of the speed-asymmetric collinear pusher considered was
approximately $0.2\%$ of the body frame beat amplitude, in contrast to $10\%$
for the Najafi-Golestanian swimmer.

The viscosity differential, rheologically-enhanced flow decay and surface
gradients of viscosity provide insight into the effects of shear-thinning on
microswimmers. While idealized, our models show that shear-thinning has both
significant and subtle effects on the trajectories and speeds of migratory
cells, emphasizing the need to take such properties into account when
investigating the physics of microswimming in complex fluids.


\subsection*{Acknowledgements}
\label{subsec:acknowledgements}

TDMJ is funded by Engineering and Physical Sciences Research Council First Grant
EP/K007637/1 to DJS. A portion of this work was completed while TDMJ was funded
by an EPSRC Doctoral Training Studentship and DJS by a Birmingham Science City
Fellowship.  Micrograph 11(b) was taken in collaboration with Dr Hermes
Gad\^{e}lha and Dr Jackson Kirkman-Brown, and micrograph 8(b) was taken by Prof.
Raymond E.  Goldstein, University of Cambridge. The authors would like to thank
Prof. John Blake for discussions and mentorship. We also acknowledge the
anonymous referees for their valuable suggestions.


\begin{appendix}
\section{A validation of the method of femlets}
\label{sec:appendix}

To validate the method of femlets, we will begin by comparing the flow arising
from an isolated, two-dimensional blob force in an enclosed circular domain of
Newtonian fluid as calculated by: (i) the method of femlets, (ii) the
established method of regularized stokeslets \citep{cortez2001method}.  For a
cut-off function of the form,
\begin{equation}
g^\epsilon(\mathbf{x}) = \frac{3\epsilon^3}{2\pi(|\mathbf{x}|^2 +
\epsilon^2)^{5/2}},
\label{eq:reg_stokeslet_cutoff}
\end{equation}
the fluid flow field arising from a single regularized stokeslet
$g^\epsilon(\mathbf{x} - \mathbf{x}_k)\mathbf{f}_k$ located at $\mathbf{x}_k$ is
given by,
\begin{align}
\mathbf{u}(\mathbf{x}) =& \frac{-\mathbf{f}_k}{4\pi\mu}\left[ \ln
\left(\sqrt{r_k^2 + \epsilon^2} + \epsilon \right) -
\frac{\epsilon\left(\sqrt{r_k^2 + \epsilon^2} + 2\epsilon
\right)}{\left(\sqrt{r_k^2 + \epsilon^2} + \epsilon \right)\sqrt{r_k^2 +
\epsilon^2}} \right] \nonumber \\
&+ \frac{1}{4\pi\mu}[\mathbf{f}_k\cdot(\mathbf{x} - \mathbf{x}_k)](\mathbf{x} -
\mathbf{x}_k)\left[ \frac{\sqrt{r_k^2 + \epsilon^2} +
2\epsilon}{\left(\sqrt{r_k^2 + \epsilon^2} + \epsilon \right)^2\sqrt{r_k^2 +
\epsilon^2}}  \right], \nonumber \\
=&\ \mathbf{S}^\epsilon(\mathbf{x},\mathbf{x}_k)\cdot\mathbf{f}_k,
\end{align}
for $r_k = |\mathbf{x} - \mathbf{x}_k|$. The outer boundary $\partial D$ is
given by $r = 10$, and a single regularized stokeslet is placed at the origin.
The flow field in domain $D$ is then given by
\begin{equation}
\mathbf{u}(\mathbf{x}) = \oint_{\partial D}
\mathbf{S}^{\epsilon_1}(\mathbf{x},\boldsymbol{\xi}(s))\cdot\mathbf{f}(s)\,\mathrm{d}s
+ \mathbf{S}^{\epsilon_2}(\mathbf{x},\mathbf{0})\cdot\mathbf{f}_{0}.
\end{equation}
for $\boldsymbol{\xi}(s)$ a parameterization of the boundary in terms of
arclength $s$. The outer boundary is discretized by $60$ equal length, constant
force elements \citep{smith2009boundary}, which correspond to the edge elements
of the finite element mesh. Each element comprises $210$ quadrature points, the
force per unit length exerted by each element on the fluid is constant, and the
regularization of the boundary stokeslets ${\epsilon_1} = 0.001$. The outer
boundary is given the no-slip velocity condition $\mathbf{u}_{\mathbf{dir}} =
\mathbf{0}$. A single regularized stokeslet with ${\epsilon_2} = 0.1$ is placed
at the origin, where the velocity is specified to be $\mathbf{u} = (1,0)$,
giving a total of $61$ degrees of freedom.

Calculating the fluid flow in the domain with the method of regularized
stokeslets is a two-stage process. Firstly, forces are calculated by specifying
velocities for each element and the central stokeslet and inverting a matrix
system. Then, these forces are used to calculate the flow at each point in the
finite element mesh. In contrast, the method of femlets calculates the forces
and flow simulataneously, and thus entails $7042$ degrees of freedom for this
example. Here, we implement the method of femlets with the same regularized
stokeslet cut-off function \eqref{eq:reg_stokeslet_cutoff}, and Dirichlet
conditions are specified on the outer boundary.

\begin{figure}[tbp]
\begin{center}
\includegraphics[scale = 0.83333]{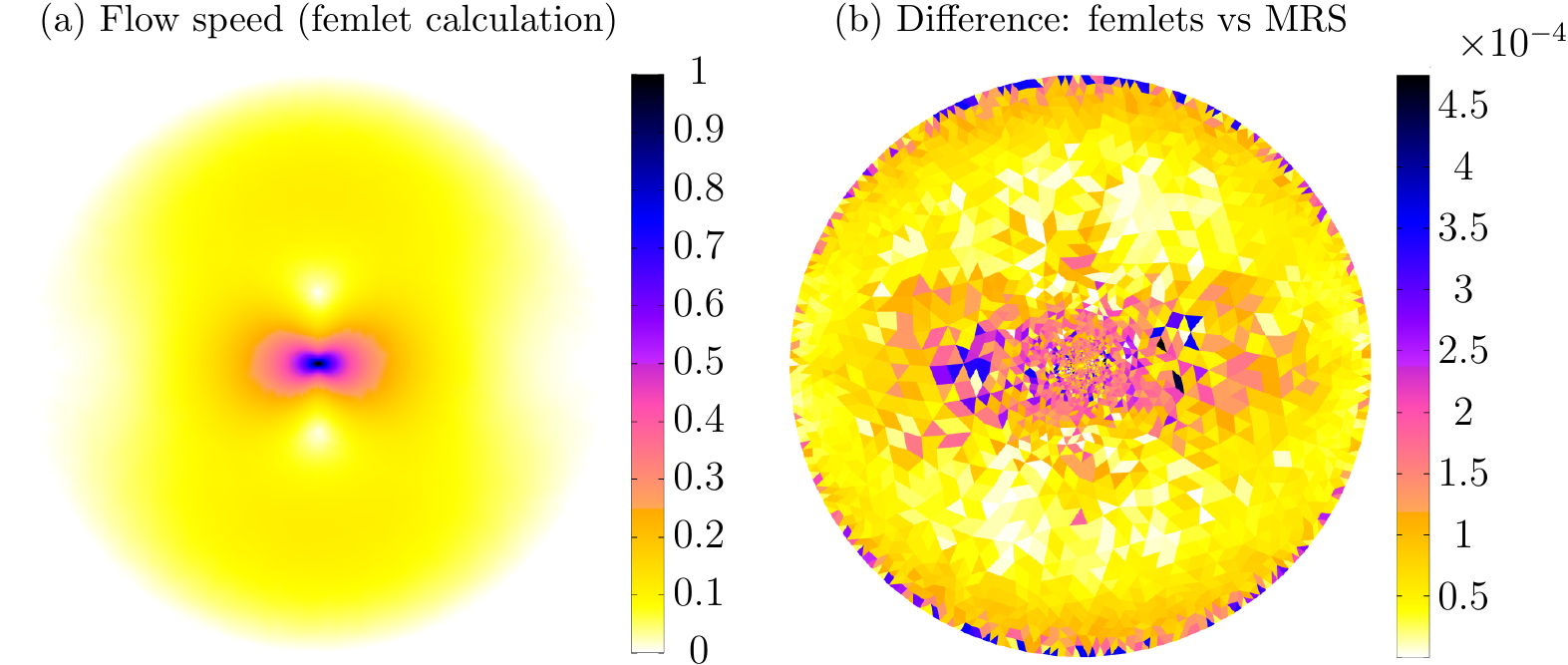}
\caption{(a) The speed of the flow arising from a regularized force of the form
\eqref{eq:reg_stokeslet_cutoff}, with $\epsilon = 0.1$, situated at the origin
in a no-slip circular cavity of radius $10$ as calculated by the method of
femlets and (b) the absolute difference between the flow speed as calculated by
the method of femlets and the method of regularized stokeslets.}	
\label{fig:isolated_force}
\end{center}
\end{figure}

Figure \ref{fig:isolated_force}a shows the speed of the flow driven by the
immersed force over the whole domain as calculated by the method of femlets,
while figure \ref{fig:isolated_force}b shows the absolute difference between the
femlet and regularized stokeslet calculations of the speed as evaluated at the
finite element mesh points. The difference is $\mathcal{O}(10^{-4})$, which is
within acceptable accuracy. Hence we conclude that the method of femlets
satisfactorily calculates the forces required to drive a specified flow.

We also wish to check that as the regularization of femlets is decreased, the
femlet solution converges to that of an equivalent moving boundary. For the
two-dimensional treadmilling squirmer of radius $r = L/2$ with slip velocity
$u_\theta = A\sin \theta$ on $r = L/2$, in infinite fluid, the swimming velocity
is given by $U = A/2$ \citep{blake1971self}. Whilst the finite element method is
only applicable for finite domains, by taking a large enough open channel we may
closely approximate a free swimmer in an infinite domain.  For a channel of
length $20L$ and height $10L$, the treadmilling squirmer is modeled by femlets
with a Gaussian cut-off function, and the regularization parameters
$\sigma_x,\sigma_y$ varied.

The calculated swimming velocity in Newtonian fluid is given as a function of
the regularizing parameters $\sigma_x,\sigma_y$ in table
\ref{tab:app_sigma_squirmer}. These results show that the difference associated
with approximating a moving boundary by femlets decreases linearly with both
$\sigma_x$ and $\sigma_y$.

\begin{table}
\begin{center}
\begin{tabular}{@{}cccccc@{}} \toprule
\multicolumn{6}{c}{Squirmer speed} \\
\# femlets & $\sigma_x$ & $\sigma_y$ & Velocity  & Rel. error &
$\mathrm{err}/\sigma_y$ \\ \midrule
$100$ & $0.0222$  & $0.0111$   & $0.25897$  & $0.0359$  & $3.23$  \\	
$100$ & $0.0222$  & $0.00555$  & $0.25462$  & $0.0185$  & $3.33$  \\
$100$ & $0.0222$  & $0.00278$  & $0.25246$  & $0.00984$ & $3.54$ \\ 
$50$  & $0.0444$  & $0.0111$   & $0.26067$  & $0.0427$  & $3.85$ \\
$200$ & $0.0111$  & $0.00555$  & $0.25421$  & $0.0168$  & $3.03$ \\
$400$ & $0.00555$ & $0.00278$  & $0.25192$  & $0.00768$ & $2.76$ \\\bottomrule
\end{tabular}
\end{center}
\caption{The velocity of the treadmilling squirmer as
calculated with the method of femlets as a function of the regularization
parameters $\sigma_x,\sigma_y$, showing that the error associated in
approximating a moving Dirichlet boundary by femlets decreases as approximately
$\mathcal{O}(\sigma_y)$.}
\label{tab:app_sigma_squirmer}
\end{table}

\begin{figure}[tbp]
\begin{center}
\includegraphics[scale = 0.83333]{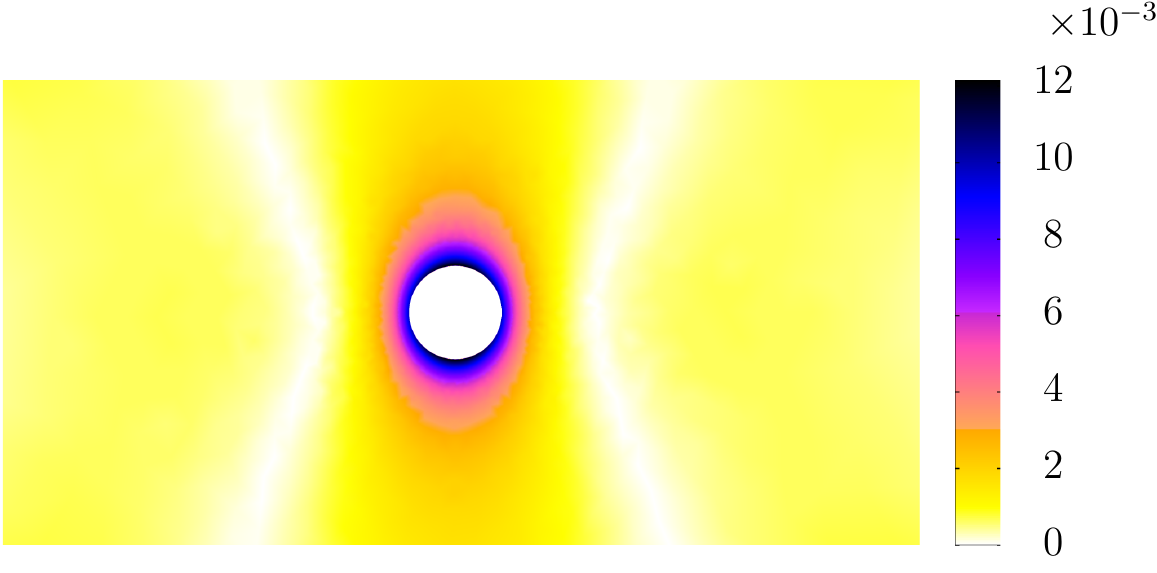}
\caption{Relative error in the calculated speed of the flow induced by the
treadmilling squirmer in Newtonian fluid, compared with the analytical solution
of \citet{blake1971self} for an infinite fluid. The maximum relative error close
to the squirmer is $1.2\%$, and is approximately $0.2\%$ throughout the majority
of the domain.}
\label{fig:squirmer_speed_error}
\end{center}
\end{figure}

The velocity field driven by the treadmilling squirmer in infinite fluid is
given in cylindrical polar coordinates by \citep{blake1971self},
\begin{subequations}
\begin{align}
u_r(r,\theta) =& \frac{1}{2}A\frac{(L/2)^2}{r^2}\cos \theta, \\
u_\theta(r,\theta) =& \frac{1}{2}A\frac{(L/2)^2}{r^2}\sin \theta. 
\label{eq:squirmer_2d_vel_field}
\end{align}
\end{subequations}
The relative error in the numerically calculated flow speed for $\sigma_x =
0.0222,\sigma_y = 0.00278$ is shown in figure \ref{fig:squirmer_speed_error}.
The error is approximately $\mathcal{O}(\sigma_y)$, and is largest in the near
field where the approximation of the boundary as an immersed regularized force
driving the flow is most apparent.

\end{appendix}


\end{document}